\documentclass[10pt,oneside,onecolumn,aps,pra,preprintnumbers,bibnotes10pt,superscriptaddress,amsmath,amssymb]{revtex4-2}

\usepackage{graphicx}
\usepackage{float}
\usepackage{braket}
\usepackage{bbold}
\usepackage{amsmath}
\usepackage{amsthm}
\usepackage[dvipsnames]{xcolor}
\usepackage{multirow}
\usepackage{tikz}
\usetikzlibrary{positioning}
\usetikzlibrary{decorations.pathmorphing, decorations.pathreplacing}
\usepackage{ragged2e}
\usepackage{commath}
\usepackage{comment}
\usepackage{adjustbox}
\usepackage{subfig}
\usepackage{hyperref}
\usepackage[normalem]{ulem}
\usepackage{figchild}

\begin{document}

\title{Experimental challenges and prospects for quantum-enhanced energy conversion: Stationary Fano coherence in V-type qutrits interacting with polarized incoherent radiation}

\author{Ludovica Donati}
\affiliation{Istituto Nazionale di Ottica, Consiglio Nazionale delle Ricerche (CNR-INO), Largo Enrico Fermi 6, 50125, Firenze, Italy.}
\affiliation{European Laboratory for Non-linear Spectroscopy (LENS), Università degli Studi di Firenze, via Nello Carrara 1, 50019, Sesto Fiorentino (FI), Italy.}

\author{Francesco Saverio Cataliotti}
\affiliation{Dipartimento di Fisica e Astronomia, Università degli Studi di Firenze, via Sansone 1, 50019, Sesto Fiorentino (FI), Italy.}
\affiliation{Istituto dei Sistemi Complessi, Consiglio Nazionale delle Ricerche (CNR-ISC), Via Madonna del Piano 10, 50019, Sesto Fiorentino (FI), Italy.}
\affiliation{European Laboratory for Non-linear Spectroscopy (LENS), Università degli Studi di Firenze, via Nello Carrara 1, 50019, Sesto Fiorentino (FI), Italy.}

\author{Stefano Gherardini}
\affiliation{Istituto Nazionale di Ottica, Consiglio Nazionale delle Ricerche (CNR-INO), Largo Enrico Fermi 6, 50125, Firenze, Italy.}
\affiliation{European Laboratory for Non-linear Spectroscopy (LENS), Università degli Studi di Firenze, via Nello Carrara 1, 50019, Sesto Fiorentino (FI), Italy.}

\date{\today}

\begin{abstract}
\textbf{Purpose:} 
Quantum coherence, once mainly studied in atomic and optical systems, now offers potential for energy conversion technologies. It influences light absorption and emission, affecting energy conversion limits and efficiency. As a result, quantum coherence is being harnessed to boost performance in quantum heat engines, photocells, and photosynthetic-inspired platforms. Of particular interest in this context is the generation of Fano coherences, i.e., the formation of quantum coherences due to the interaction with the continuum of modes characterizing an incoherent process. We aim to formalize mathematically the possibility of achieving steady-state Fano coherence in a V-type three-level quantum system using polarized incoherent radiation, without requiring the energy difference between the excited levels, $\hbar\Delta$, to tend to zero. Specifically, in this scenario, it can be shown that the maximum steady-state Fano coherence is obtained through a non-trivial interplay between $\Delta$, the average spontaneous decay rate $\bar{\gamma}$ of the excited levels, and the intensity of the incoherent radiation, quantified by the mean photon number $\bar{n}$.\\
\indent\textbf{Methods:} 
We perform this analysis by deriving the Bloch-Redfield equation from first-principles by quantizing the incoherent radiation. The resulting reduced dynamics of the system are analysed, so as to determine the lifetime of Fano coherence and identify the conditions under which it becomes stationary.\\ 
\indent\textbf{Results:} 
We characterise distinct dynamical regimes, ranging from weak to strong pumping, in which steady-state Fano coherence emerges, and we quantitatively determine its magnitude. For each regime, we analyse the generation of Fano coherence as a function of both the intensity of the incoherent pumping and the energy splitting between the excited levels.
We also assess how obtaining Fano coherence is modified by symmetric or asymmetric decay rates. These findings indicate that a three-level quantum system driven by polarized incoherent light can act as a robust resource for coherence-assisted energy conversion and storage. Finally, we discuss the experimental challenges associated with the implementation of the proposed model using an ensemble of Rubidium atoms.\\
\indent\textbf{Conclusion:} 
This work paves the way towards the forthcoming realization of experimental tests of Fano coherence generation in a relevant experimental setup by making use of a polarized incoherent radiation, in order to achieve quantum-enhanced energy storage and energy-conversion functionalities.
\end{abstract}

\maketitle

\tableofcontents

%%%%%%%%%%%%%%%%%%%%%%%%%%%%%%%%%%%%%%
\section{Introduction}\label{sec1}

Quantum coherence, traditionally investigated in atomic and optical systems, is a promising resource for energy conversion technologies. It can indeed alter the balance between absorption and emission processes of light by matter, which traditionally bounds energy conversion performance and enter the definition of entropy production, work extraction and transport fluxes. Therefore, mechanisms exploiting quantum coherence have been proposed to enhance output power and efficiency in driven quantum systems, quantum heat engines and photocells, as well as to improve energy-transfer yields in biomimetic and photosynthetic-inspired architectures~\cite{Scully-Chapin,Um,Scully,Svidzinsky:Scully,Wang,Lira,Creatore,Svidzinsky,Dorfman,Bittner,Tomasi,CampaioliRMP2024,GherardiniPRXQuantum2024}.

Quantum heat engines (QHEs) provide a striking example about using quantum coherence as a thermodynamic resource. Primarily, quantum coherence can be externally induced using a coherent driving field that creates superpositions between excited energy levels~\cite{Scully-Chapin}. Such a drive has the effect to increase the extractable work, thus providing a substantial enhancement of the engine performance. Beyond the coherent interaction with an external field, quantum coherence can arise from the Hamiltonian coupling of two quantum systems; for example, solid-state platforms based on degenerate double quantum dots, coupled in parallel to two thermal reservoirs~\cite{Um}. In this configuration, coherent tunnelling between quantum dots generates steady-state quantum coherence in their state. 

Concerning solar conversion devices, quantum coherence has been identified as a mechanism capable of reducing radiative recombination losses and improving photovoltaic performance. In particular, early theoretical proposals introduced quantum-dot-based photocells driven by a monochromatic component of the solar spectrum, establishing the foundation for coherence-assisted operation and predicting significant gains in conversion efficiency~\cite{Scully,Svidzinsky:Scully}.
A complementary strategy was explored in Ref.~\cite{Wang}, where optimal tunneling between two quantum dots was shown to sustain quantum coherence between localized states, enhancing the quantum photovoltaic effect and increasing current and power output under incoherent illumination.  More recently, quantum-dot molecules has been investigated, where coherent tunneling through a finite barrier generates steady-state quantum coherence that, under Markovian master equation dynamics, can improve the efficiency at maximum power by up to 30\%~\cite{Lira}. Inspired by photosynthetic reaction centers, Creatore et al. have introduced in Ref.~\cite{Creatore} a nanoscale photocell model composed of molecular chromophores showing that dipole-dipole mediated quantum interference in absorption and emission processes can increase photocurrent and power output by up to 35\% compared to a photovoltaic cell lacking such coherent effects.

Remarkably, quantum coherence can emerge even when a system interacts with an incoherent environment, such as a thermal or vacuum reservoir~\cite{Agarwal1,Hegerfeldt}. In particular, as discussed in Refs.~\cite{Latune,Latune2}, it comes from the indistinguishability of the decay or excitation pathways for the system's dynamics due to the interaction with the environment. 
In energy conversion contexts,  this effect, also referred to as \emph{noise-induced coherence}, justifies the enhancement of photocurrent and power output in quantum-dot-based photocells without requiring external coherent driving~\cite{Svidzinsky,Scully-Chapin}, as well as the enhancement of energy transfer efficiency in biologically inspired systems under incoherent solar pumping~\cite{Dorfman}. Experimentally, in polymeric semiconductor heterojunctions, noise-induced coherence drives ultrafast photo-carrier generation dynamics, underscoring its role in real photovoltaic materials~\cite{Bittner}. A recent theoretical work has further demonstrated that environmental interactions do not necessarily degrade coherent advantages; on the contrary, a strong system reservoir coupling can improve conversion efficiency thanks to optically induced coherence, broadening the range of platforms where coherent light harvesting can be realized~\cite{Tomasi}.

All the aforementioned platforms can be effectively modelled as multi-level quantum systems interacting with an external environment. The minimal setting in which quantum interference can arise is a three-level system. In this work, we focus on a V-type three-level system, consisting of a common ground level and two nearly degenerate excited levels, each optically coupled to the ground through interaction with an incoherent radiation field, such as broadband thermal or solar radiation.
Incoherent pumping of a V-type system with nearly degenerate excited levels can give rise to noise-induced quantum coherence between the excited states \cite{Tscherbul-Brumer2,Tscherbul-Brumer,Dodin-Tscherbul}. This effect originates from Fano interference, which occurs when multiple discrete transitions couple to a common continuum of environmental modes~\cite{Fano,Fano2}. In the present context, the continuum is provided by the broadband spectrum of the radiation field or, equivalently, by the effective thermal reservoir with which the system exchanges energy.
When the spectral width of the incoherent radiation exceeds the frequency splitting of the excited levels, the excitation and decay pathways connecting these states to the ground state become indistinguishable. This indistinguishability enables quantum interference between the corresponding optical transitions, leading to the formation of noise-induced Fano coherence, even in the absence of external coherent driving.

Recent works have extensively examined the dynamics of multi-level quantum systems, particularly V-type three-level systems, driven by isotropic, unpolarized, incoherent radiation~\cite{Kozlov,Tscherbul-Brumer2,Tscherbul-Brumer,Dodin-Tscherbul,Dodin-Tscherbul2,Koyu-Tscherbul, Donati}. These studies identify distinct dynamical regimes governed by the ratio between the frequency splitting of excited levels, $\Delta$, and their average spontaneous decay rate $\bar{\gamma}$. The system's dynamics are further influenced by the mean photon number $\bar{n}$ of the incident radiation, which, in these models, is determined by the ratio between the average incoherent excitation rate $\bar{r}$ and $\bar{\gamma}$~\cite{Koyu-Tscherbul,Dodin-Tscherbul,Tscherbul-Brumer2,Donati}.

Much less explored is the case of an anisotropic, polarized, incoherent radiation as the input source~\cite{Dodin-Brumer,Koyu-Dodin}. Polarization and anisotropy introduce an additional level of complexity in the system environment interaction, by selecting specific modes of the field and coupling directions. Unlike isotropic unpolarized radiation, an anisotropic polarized field does not average over all propagation directions, thus preserving phase correlations between optical transitions that sustain quantum interference effects. This mechanism enables the generation and stabilization of noise-induced Fano coherence between excited levels, even when quantum interference would otherwise vanish under isotropic driving.

In this manuscript, we provide a first-principle quantum modelling for the dynamics of a V-type three-level system driven by anisotropic polarized incoherent radiation. In particular, we analyse how this form of pumping leads to the emergence of stationary noise-induced Fano coherence across all dynamical regimes. Particular attention is devoted to identifying the conditions that maximize the steady-state coherence, with a view toward its experimental detection in atomic systems. Despite extensive theoretical investigations and clear technological relevance, experimental evidence of Fano coherences generated by incoherent pumping is still missing, to the best of our knowledge.

The paper is organized as follows. In Sec.~\ref{sec2}, we introduce the theoretical model by deriving the open quantum system dynamics of a V-type three-level system interacting with an incoherent radiation. Subsec.~\ref{subsec2.1} presents the system field interaction Hamiltonian, while Subsec.~\ref{subsec2.2} outlines the Bloch–Redfield formalism used to describe the reduced dynamics of the system. Within this framework, we discuss all the approximations we need to derive a Markovian master equation, and we carefully distinguish between isotropic and anisotropic incoherent radiation fields. The resulting dynamical equations are cast in a linear form suitable for analytical and numerical analysis. In Subsec.~\ref{subsec:2.3}, we analyze the system dynamics across different regimes, identifying weak- and strong-pumping conditions and examining the lifetime and behaviour of Fano coherences. Sec.~\ref{sec3} focuses on experimentally relevant implementations. In particular, in Subsec.~\ref{subsec:3.1}, we explain our choice of an ensemble of $^{87}$Rb atoms as the physical platform, and in Subsec.~\ref{subsec:3.2}, we describe how a V-type three-level system can be realized within its hyperfine structure. We then present quantitative theoretical predictions for the resulting steady-state Fano coherence and identify the parameter regimes that are most favourable for experimental observation.
Finally, in Sec.~\ref{sec4} we summarize our main results, discussing possible future research directions in Subsec.~\ref{subsec:4.1}.

%%%%%%%%%%%%%%%%%%%%%%%%%%%%%%%%

\section{Model}\label{sec2}

\subsection{System-incoherent light interaction Hamiltonian}\label{subsec2.1}

Let us consider a V-type three-level quantum system driven by an incoherent radiation source, such as thermal or broadband radiation. This scenario is modelled using the formalism of open quantum systems~\cite{Petruccione}, which assumes that the system under analysis (S) interacts with an external reservoir (R) with possibly macroscopic features. The resulting composite system S$+$R reservoir is considered as a closed system.

\begin{figure}[t]
\centering
\includegraphics[width=0.5\textwidth]{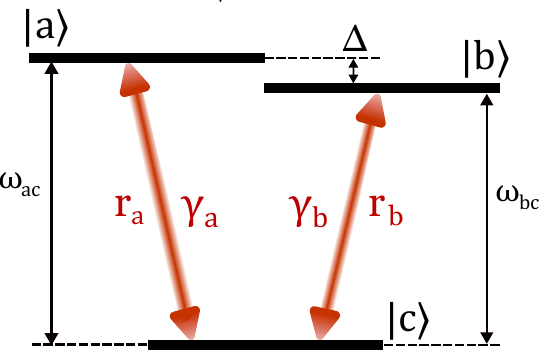}
\caption{
Pictorial representation of the V-type three-level system considered in our analysis. Its energy level configuration consists of two nearly degenerate excited levels, denoted as $|a\rangle$ and $|b\rangle$ with a frequency splitting of $\rm \Delta$. These levels are incoherently pumped, at rates ${\rm r}_a$ and ${\rm r}_b$ respectively, from the ground level $|c\rangle$. Both $|a\rangle$ and $|b\rangle$ can decay to the ground level at rates $\gamma_a$ and $\gamma_b$.}
\label{fig:1}
\end{figure}

The configuration of the V-type three-level system consists of two excited states and a single ground state, as shown in Fig.~\ref{fig:1}. The excited levels $|a\rangle$ and $|b\rangle$ decay to the ground state $|c\rangle$ with spontaneous emission rates $\gamma_a$ and $\gamma_b$, respectively. Additionally, both excited states are coupled to the ground via incoherent pumping (thermal radiation in our model) with rate ${\rm r}_a, {\rm r}_b$. There is no direct dipole moment coupling between the two upper levels.
The angular transition frequencies are given by $\omega_{ac} = \omega_a - \omega_c$ and $\omega_{bc} = \omega_b - \omega_c$, with their difference $\Delta = \omega_{ac} - \omega_{bc}$ defining the frequency splitting between the excited levels.
The two excited states are nearly degenerate, meaning that the splitting $\Delta$ is much smaller than the optical transition frequencies ($\Delta \ll \omega_{ac}, \omega_{bc}$).  
In this way, the level separation becomes comparable to the characteristic energy scale of the system--reservoir interaction. As a result, the system dynamics may exhibit pronounced reservoir-induced interference effects, as reported in several previous studies~\cite{Tscherbul-Brumer2,Tscherbul-Brumer,Li,Dodin-Tscherbul,Dodin-Tscherbul2,Koyu-Tscherbul,Ivander}.
In the model we are going to derive, the electromagnetic field acts as a thermal reservoir, and the dynamics of the combined three-level system $+$ incoherent radiation are treated entirely within a quantum mechanical description.

The time-independent Hamiltonian of the reduced system S is expressed as:
\begin{equation}\label{eq:1.1}
\hat{H}_{S} = \sum_l\hbar\omega_{lc}\hat{\sigma}_{lc}^+\hat{\sigma}_{lc}^-=\hbar\left(\omega_{ac}\hat{\sigma}_{ac}^+\hat{\sigma}_{ac}^- + \omega_{bc}\hat{\sigma}_{bc}^+\hat{\sigma}_{bc}^-\right),
\end{equation}
where $l=a,b$, and, without loss of generality, we have set the energy of the ground level $|c\rangle$ to zero. 
In Eq.~\eqref{eq:1.1}, $\hat{\sigma}_{lc}^-$ and $\hat{\sigma}_{lc}^+$ are the lowering and raising operators, respectively, which are defined as:
\begin{equation}
  \hat{\sigma}_{lc}^{+} \equiv |l\rangle\langle c| \quad \text{and} \quad \hat{\sigma}_{lc}^- \equiv \left(\hat{\sigma}_{lc}^+\right)^\dagger = |c\rangle\langle l|.
\end{equation}  
As $|l\rangle$ is an excited state and $|c\rangle$ is the ground state, $\hat{\sigma}_{lc}^+$ represents the transition from the ground to the excited state (raising), and $\hat{\sigma}_{lc}^-$ denotes the transition from the excited to the ground state (lowering).
Equivalently, the Hamiltonian of the quantized electromagnetic field is
\begin{equation}\label{eq:1.3}
\hat{H}_R =  \sum_{\lambda=1,2}\sum_{\bf k} \hbar\nu_{{\bf k},\lambda}\hat{a}_{{\bf k},\lambda}^\dagger\hat{a}_{{\bf k},\lambda},
\end{equation}
where $\nu_{{\bf k},\lambda}$ is the angular frequency of the mode with wavevector $\bf k$ and polarization $\lambda$. The operators $\hat{a}_{{\bf k},\lambda}$ and $\hat{a}_{{\bf k},\lambda}^\dagger$ annihilate and create, respectively, a photon in the corresponding mode. The Hamiltonian $\hat{H}_R$ represents the total energy stored in all field modes, provided the energy of the vacuum $\tfrac{1}{2}\hbar \nu_{\mathbf{k},\lambda}$ is omitted. The vacuum energy is usually discarded because it adds only an overall constant shift to the energy and has no effect on observable dynamics, which depend solely on energy differences.
For the discussion that will follow about modelling the interaction with an anisotropic incoherent radiation, the modes of the reservoir could emit and absorb (independently) photons sampled from thermal distributions at different temperatures, depending on the amount of anisotropy. In particular, to clarify, if the incoherent radiation field is isotropic, the absorbed and emitted photons are sampled from a single thermal distribution at finite temperature. Otherwise, we need to consider the effective interaction of the system with two thermal reservoirs at different temperatures: an anisotropic, polarized radiation field at finite temperature, responsible for absorption and stimulated emission, and an isotropic radiation field modelled as a zero-temperature reservoir, accounting exclusively for spontaneous emission.

The Hamiltonian of the composite system S+R includes the interaction Hamiltonian $\hat{H}(t)$, which describes the interaction between the system and the electromagnetic field. The Hamiltonian $\hat{H}(t)$ can be derived by considering the coupling between the electric dipole-moment operator $\hat{\mu}$ of the V-system, and the electric field operator $\hat{E}({\bf r},t)$~\cite{Scully:Zub,Steck,Petruccione}. Before proceeding, notice that throughout the paper, operators that are not labelled with the subscript $I$ denote quantities expressed in the Schr\"{o}dinger picture. In the interaction picture, denoted by the subscript $I$, the following expression is obtained:
\begin{equation}\label{eq:1.4}
\hat{H}_{I}(t)=-\hat{\mu}_{I}(t)\cdot\hat{E}_{I}({\bf R},t),
\end{equation}
where 
\begin{equation}\label{eq:1.5}
\hat{E}_I({\bf R},t)=i\sum_{\lambda=1,2}\sum_{\bf k}\sqrt{\frac{\hbar\nu_{{\bf k},\lambda}}{2\varepsilon_0V}}{\bm\epsilon}_{{\bf k},\lambda}\left[\hat{a}_{{\bf k},\lambda}e^{-i\nu_{{\bf k},\lambda}t + i{\bf k}\cdot {\bf R}} - \hat{a}_{{\bf k},\lambda}^\dagger e^{i\nu_{{\bf k},\lambda}t - i{\bf k}\cdot {\bf R}} \right],
\end{equation}
with the normalization prefactor resulting from field quantization in a volume $V$.
In~Eq. \eqref{eq:1.5}, ${\bm\epsilon}_{{\bf k},\lambda}$ is the unit polarization vector, which must satisfy the transversality condition imposed by the Coulomb gauge:
\begin{equation}\label{eq:1.6}
{\bm\epsilon}_{{\bf k},\lambda} \cdot {\bf k} = 0 \quad \forall\, {\bf k},\lambda .
\end{equation} 
Condition \eqref{eq:1.6} assures the transversality of the electromagnetic field, excluding any longitudinal polarization component along the direction the electromagnetic wave travels. This means that the electric field lies in the plane perpendicular to the wavevector $\bf k$, admitting two independent orthonormal transverse polarization vectors ${\bm\epsilon}_{{\bf k},1}$ and ${\bm\epsilon}_{{\bf k},2}$. Any polarization state associated with a given wavevector ${\bf k}$ can be expressed as a linear combination of these two transverse modes. Their orthonormality is expressed as
\begin{equation}\label{eq:1.7}
{\bm\epsilon}_{{\bf k},\lambda} \cdot {\bm\epsilon}_{{\bf k}',\lambda'} = \delta_{\lambda,\lambda'} \quad \forall\, {\bf k},
\end{equation} 
with here $\delta$ denoting the Kronecker delta.

In the interaction picture, the dipole moment operator for the V-type three level system shown in Fig.~\ref{fig:1} is:
\begin{equation}\label{eq:1.8}
\hat{\mu}_{I}({\bf R}, t) = {\bm\mu}_{ac}\left(\hat{\sigma}_{ac}^+e^{i\omega_{ac}t} +\hat{\sigma}_{ac}^-e^{-i\omega_{ac}t}\right) + {\bm\mu}_{bc}\left(\hat{\sigma}_{bc}^+e^{i\omega_{bc}t}+\hat{\sigma}_{bc}^-e^{-i\omega_{bc}t}\right),
\end{equation}
where ${\bm\mu}_{ac}\equiv\langle a|e\bf R|c\rangle$ and ${\bm\mu}_{bc}\equiv\langle b|e\bf R|c\rangle$ are the matrix elements of the electric dipole moment operator, $e$ is the elementary charge, and $\bf R$ is the position vector.
Eq.~\eqref{eq:1.8} follows from the canonical transformation that maps the Schr\"{o}dinger picture to the interaction picture, namely 
\begin{equation}\label{eq:1.9}
\hat{\sigma}_{lc,I}^{\pm}=\hat{U}_{0}^\dagger(t,t_0)\hat{\sigma}_{lc}^\pm\hat{U}_{0}(t,t_0)=\hat{\sigma}_{lc}^\pm e^{\pm i\omega_{lc}(t-t_0)},
\end{equation}
where
\begin{equation}\label{eq:1.10}
  \hat{U}_{0}(t,t_0) \equiv \exp\left[-i\sum_{l}\omega_{lc}\hat{\sigma}_{ lc}^+\hat{\sigma}_{lc}^-(t-t_0)\right],
\end{equation}
which is a unitary time-evolution operator defined with respect to the initial time $t_0$.

Inserting Eqs.~\eqref{eq:1.5} and \eqref{eq:1.8}, where we assume $t_0=0$, in Eq.~\eqref{eq:1.4} yields the following:
\begin{equation}\label{eq:1.11}
\begin{aligned}
&\hat{H}_{I}(t)=-i\sum_{\lambda=1,2}\sum_{\bf k}\sqrt{\frac{\hbar\nu_{{\bf k},\lambda}}{2\varepsilon_0V}}\Big[\left({\bm\mu}_{ac}\cdot{\bm\epsilon}_{{\bf k},\lambda}\right)\left(\hat{\sigma}_{ac}^+e^{i\omega_{ac}t} + \hat{\sigma}_{ac}^-e^{-i\omega_{ac}t} \right)\\
& \times \left(\hat{a}_{{\bf k},\lambda}e^{-i\nu_{\bf k,\lambda}t + i{\bf k}\cdot {\bf R}} - \hat{a}_{{\bf k},\lambda}^\dagger e^{i\nu_{\bf k,\lambda}t - i{\bf k}\cdot {\bf R}} \right)\Big] \\
&- i\sum_{\lambda=1,2}\sum_{\bf k}\sqrt{\frac{\hbar\nu_{{\bf k},\lambda}}{2\varepsilon_0V}}\Big[ \left({\bm\mu}_{bc}\cdot{\bm\epsilon}_{{\bf k},\lambda}\right)\left(\hat{\sigma}_{bc}^+e^{i\omega_{bc}t} + \hat{\sigma}_{bc}^-e^{-i\omega_{bc}t} \right)\\
& \times\left(\hat{a}_{{\bf k},\lambda}e^{-i\nu_{\bf k,\lambda}t + i{\bf k}\cdot {\bf R}} - \hat{a}_{{\bf k},\lambda}^\dagger e^{i\nu_{\bf k,\lambda}t - i{\bf k}\cdot {\bf R}} \right)\Big].
\end{aligned}
\end{equation}
At this point, we apply the \emph{dipole approximation}.

\begin{itemize}
\item {\bf Dipole approximation:}\\
The dipole approximation assumes that the wavelength of the electric field is much larger than the characteristic spatial extent of the system S. As a result, the electric field can be considered constant. The electric field operator $\hat{E}({\bf R},t)$ is thus evaluated at a fixed reference point, conventionally taken to be ${\bf R}=0$ for simplicity, leading to:

\begin{equation}\label{eq:1.12}
\hat{E}({\bf R},t)\approx\hat{E}(0,t)=i\sum_{\lambda=1,2}\sum_{\bf k}\sqrt{\frac{\hbar\nu_{{\bf k},\lambda}}{2\varepsilon_0V}}{\bm\epsilon}_{{\bf k},\lambda}\left[\hat{a}_{{\bf k},\lambda}e^{-i\nu_{\bf k,\lambda}t} - \hat{a}_{{\bf k},\lambda}^\dagger e^{i\nu_{\bf k,\lambda}t} \right]. 
\end{equation}
\end{itemize}
Hence, Eq.~\eqref{eq:1.11} becomes:
\begin{equation}\label{eq:1.13}
\begin{aligned}
&\hat{H}_I(t)=-i\sum_{\lambda=1,2}\sum_{\bf k}\sqrt{\frac{\hbar\nu_{{\bf k},\lambda}}{2\varepsilon_0V}}\Big[\left({\bm\mu}_{ac}\cdot{\bm\epsilon}_{{\bf k},\lambda}\right)\left(\hat{\sigma}_{ac}^+e^{i\omega_{ac}t} + \hat{\sigma}_{ac}^-e^{-i\omega_{ac}t} \right)\left(\hat{a}_{{\bf k},\lambda}e^{-i\nu_{\bf k,\lambda}t} - \hat{a}_{{\bf k},\lambda}^\dagger e^{i\nu_{\bf k,\lambda}t} \right)\Big] \\
&- i\sum_{\lambda=1,2}\sum_{\bf k}\sqrt{\frac{\hbar\nu_{{\bf k},\lambda}}{2\varepsilon_0V}}\left[ \left({\bm\mu}_{bc}\cdot{\bm\epsilon}_{{\bf k},\lambda}\right)\left(\hat{\sigma}_{bc}^+e^{i\omega_{bc}t} + \hat{\sigma}_{bc}^-e^{-i\omega_{bc}t} \right)\left(\hat{a}_{{\bf k},\lambda}e^{-i\nu_{\bf k,\lambda}t} - \hat{a}_{{\bf k},\lambda}^\dagger e^{i\nu_{\bf k,\lambda}t} \right)\right] .
\end{aligned}
\end{equation}

Eq.~\eqref{eq:1.13} can be further simplified by applying the Rotating Wave Approximation (RWA) that allows to discard the non-conserving energy terms contained in the equation, as described below.

\begin{itemize}
\item 
{\bf Rotating Wave Approximation:}\\ Considering the Eq.~\eqref{eq:1.13}, not all the four terms that arise from expanding the product 
$$
\left( \hat{\sigma}_{lc}^+e^{i\omega_{lc}} + \hat{\sigma}_{lc}^-e^{-i\omega_{lc}} \right)\left( \hat{a}_{{\bf k},\lambda}e^{-i\nu_{\bf k}t} - \hat{a}_{{\bf k},\lambda}^\dagger e^{i\nu_{\bf k}t} \right)
$$, with $l=a,b$, conserve energy. Specifically, the term $\hat{\sigma}_{lc}^-\hat{a}_{{\bf k},\lambda}$ describes the transition from the excited state $|l\rangle$ to the ground state $|c\rangle$ of the system, and the destruction of one photon in the field. This results in a net energy loss, equal to the sum of the photon energy and the energy difference between the states of the system. Conversely, the term $\hat{\sigma}_{lc}^+\hat{a}_{{\bf k},\lambda}^\dagger$ represents the transition of the system's state to the excited state $|l\rangle$, and the creation of one photon in the field, thus resulting in an energy gain. On the other hand, the remaining two terms conserve the energy of the light-matter interaction process~\cite{Scully:Zub, Petruccione, Steck}. RWA consists in neglecting the non-energy-conserving (or counter-rotating) terms $\hat{\sigma}_{lc}^-\hat{a}_{{\bf k},\lambda}$ and $\hat{\sigma}_{lc}^+\hat{a}_{{\bf k},\lambda}^\dagger$.  
Hence, Eq.~\eqref{eq:1.13} turns into:
\begin{equation}\label{eq:1.14}
\begin{aligned}
&\hat{H}_I(t)=-i\sum_{\lambda=1,2}\sum_{\bf k}\sqrt{\frac{\hbar\nu_{{\bf k},\lambda}}{2\varepsilon_0V}}\left[\left({\bm\mu}_{ac}\cdot{\bm\epsilon}_{{\bf k},\lambda}\right)\left(\hat{\sigma}_{ac}^+\hat{a}_{{\bf k},\lambda}e^{i\left(\omega_{ac}t-\nu_{\bf k,\lambda}t\right)} - \hat{\sigma}_{ac}^-\hat{a}_{{\bf k},\lambda}^\dagger e^{-i\left(\omega_{ac}t-\nu_{\bf k,\lambda}t\right)}  \right)\right] \\
&-i \sum_{\lambda=1,2}\sum_{\bf k}\sqrt{\frac{\hbar\nu_{{\bf k},\lambda}}{2\varepsilon_0V}}\left[ \left({\bm\mu}_{bc}\cdot{\bm\epsilon}_{{\bf k},\lambda}\right)\left(\hat{\sigma}_{bc}^+\hat{a}_{{\bf k},\lambda}e^{i\left(\omega_{bc}t-\nu_{\bf k,\lambda}t\right)} - \hat{\sigma}_{bc}^-\hat{a}_{{\bf k},\lambda}^\dagger e^{-i\left(\omega_{bc}t-\nu_{\bf k,\lambda}t\right)}\right)\right].
\end{aligned}
\end{equation}
\end{itemize}
Eq.~\eqref{eq:1.14} can be rewritten also using the \emph{coupling terms} $g_{\bf{k},\lambda}^{(l)}$, which quantify the strength of the interaction between S and R. The coupling terms have dimensions of an angular frequency:
\begin{equation}\label{eq:1.15}
g_{\bf{k},\lambda}^{(l)} \equiv -\frac{\bm{\mu}_{lc}\cdot\bm{\epsilon}_{\bf{k},\lambda}}{\hbar}\sqrt{\frac{\hbar\nu_{\bf{k},\lambda}}{2\varepsilon_0V}},    
\end{equation}
so that
\begin{equation}\label{eq:1.16}
\begin{aligned}
\hat{H}_I(t)=&i\hbar\sum_{\lambda=1,2}\sum_{\bf k}g_{\bf{k},\lambda}^{(a)}\left(\hat{\sigma}_{ac}^+\hat{a}_{{\bf k},\lambda}e^{i\left(\omega_{ac}t-\nu_{\bf k,\lambda}t\right)} - \hat{\sigma}_{ac}^-\hat{a}_{{\bf k},\lambda}^\dagger e^{-i\left(\omega_{ac}t-\nu_{\bf k,\lambda}t\right)}  \right) \\
 +& i\hbar\sum_{\lambda=1,2}\sum_{\bf k}g_{\bf{k},\lambda}^{(b)}\left(\hat{\sigma}_{bc}^+\hat{a}_{{\bf k},\lambda}e^{i\left(\omega_{bc}t-\nu_{\bf k,\lambda}t\right)} - \hat{\sigma}_{bc}^-\hat{a}_{{\bf k},\lambda}^\dagger e^{-i\left(\omega_{bc}t-\nu_{\bf k,\lambda}t\right)}\right).
\end{aligned}
\end{equation}
In Eq.~\eqref{eq:1.15}, the dipole matrix element ${\bm\mu}_{lc}=e\langle l|{\bf R}|c\rangle$ can be chosen to be real, so that ${\bm\mu}_{lc}={\bm\mu}_{cl}^{*}$~\cite{Steck}.

The coupling terms $g_{\bf{k},\lambda}^{(l)}$ define the \emph{spectral density function} $J_l(\nu)$, which characterises the distribution of the radiation field modes. The spectral density is formally given by:
\begin{equation}\label{eq:1.17}
J_l(\nu)\equiv\sum_{\lambda=1,2} \sum_{\bf{k}}|g_{\bf{k},\lambda}^{(l)}|^2\delta(\nu-\nu_{{\bf k},\lambda}),
\end{equation} 
where $\delta(\cdot)$ denotes the Dirac delta function. The presence of $\delta(\nu-\nu_{{\bf k},\lambda})$ ensures that only the modes of the radiation field that have frequencies equal to $\nu_{{\bf k},\lambda}$ are taken into account.

The contribution of the counter-rotating terms to the dynamics of the system is suppressed, and the RWA is therefore justified, when a clear separation of timescales exists between the fast oscillations associated with the sum frequencies $\omega_{lc}+\nu_{\mathbf{k},\lambda}$ ($l=a,b$) 
and the slower time evolution governing population and coherence dynamics. The latter occurs on timescales set by the strength of the system-reservoir coupling.
In practice, a clear separation of timescales requires: (i) a weak-coupling regime $g_{\mathbf{k},\lambda}^{(l)}\ll \omega_{lc}$, and (ii) an incoherent radiation field whose spectral density is centred on the system's transitions, such that the dominant modes of the field satisfy $|\omega_{lc}-\nu_{\mathbf{k},\lambda}| \ll \omega_{lc} + \nu_{\mathbf{k},\lambda}$~\cite{Scully:Zub, Petruccione, Steck}. Under these conditions, the counter-rotating terms oscillate on timescales much shorter than those associated with decoherence and relaxation, and their contributions can be averaged out upon coarse graining over the timescale of the system dynamics. As a result, the interaction dynamics are dominated by first-order absorption and emission processes, while off-resonant second-order effects, such as scattering processes and Bloch-Siegert shifts, are strongly suppressed~\cite{Fleming,Larson,Kockum}.

%%%%%%%%%%%%%%%%%%%%%%%%%%%%%%%%
\subsection{Open system dynamics: Bloch-Redfield equation}\label{subsec2.2}

\subsubsection{Some notions from open quantum systems' theory}\label{subsubsec:1}

The system S is described as an open quantum system interacting with its environment R. We describe the state of the system via a density operator formalism, as quantum coherences can be generated during the dynamics. In our setting, as shown below, the time evolution of the system's density operator $\hat{\rho}_S(t)$ is derived using a quantum Markovian master equation. The latter is obtained from the \emph{Liouville-von Neumann equation} that governs the dynamics of the density operator $\hat{\rho}(t)$ for the composite system formed by S and R. This model, as well as its solution, has been already discussed in several references so far~\cite{Kozlov,Tscherbul-Brumer, Koyu-Tscherbul,Koyu-Dodin,Svidzinsky,Dodin-Tscherbul,Dodin-Tscherbul2,Petruccione, Donati}. We report below a comprehensive and pedagogical derivation of the model.

The time evolution of the system$+$reservoir is governed by the following Liouville-von Neumann differential equation in integro-differential form for the density operator $\hat{\rho_I}(t)$ of the composite system, expressed in interaction picture: 
\begin{equation}
\frac{d\hat{\rho_I}(t)}{dt} = -\frac{i}{\hbar}\left[ \hat{H}_I(t), \hat{\rho_I}(0)
\right]-\frac{1}{\hbar^2}\int_{0}^{t}\Big[ \hat{H}_I(t), \left[ \hat{H}_I(t'), \hat{\rho_I}(t')\right] \Big] dt'.
 \label{eq:1.18}
\end{equation}
Applying the partial trace over the reservoir degrees of freedom, we formally get:
\begin{eqnarray}\label{eq:1.19}
\frac{d\hat{\rho}_{S,I}(t)}{dt} &=& -\frac{i}{\hbar}{\rm Tr}_R\left(\left[ \hat{H}_I(t), \hat{\rho}_{S,I}(0) \otimes \hat{\rho}_{R,I}(0) \right]\right) \nonumber \\
&-& \frac{1}{\hbar^2}\int_0^t {\rm Tr}_R\left( \left[ \hat{H}_I(t), \left[ \hat{H}_I(t'), \hat{\rho}_I(t')\right]\right]\right) dt',
\end{eqnarray} 
which still comprises the density operator $\hat{\rho}_I(t)$ of the composite system. Through suitable approximations relevant to the specific scenario, it is possible to derive a differential equation that depends exclusively on the state of S at the current time $t$. The used approximations are the following:    
\begin{itemize}
\item {\bf Born approximation:}\\
The \emph{Born} (or \emph{weak-coupling approximation}) assumes that the coupling between the system and reservoir is sufficiently weak that the reservoir state is only negligibly perturbed by its interaction with the system. Under this assumption, system–reservoir correlations, generated by the interaction, remain perturbative and can be neglected. This means that the density operator of the system$+$reservoir can be approximated by the tensor product: 
\begin{equation}\label{eq:1.20}
\hat{\rho}(t)\approx\hat{\rho}_S(t)\otimes\hat{\rho}_R(0) \quad \forall \, t,
\end{equation}
with the reservoir remaining approximately in its initial state $\hat{\rho}_R(0)$ during the interaction.

By applying the Born approximation, Eq.~\eqref{eq:1.19} gets the form
\begin{eqnarray}\label{eq:1.21}
\frac{d\hat{\rho}_{S,I}(t)}{dt} &\approx & -\frac{i}{\hbar}{\rm Tr}_R\left(\left[ \hat{H}_I(t), \hat{\rho}_{S,I}(0) \otimes \hat{\rho}_{R,I}(0) \right]\right) \nonumber \\
&-&\frac{1}{\hbar^2}\int_0^t {\rm Tr}_R \left( \left[ \hat{H}_I(t), \left[ \hat{H}_I(t'), \hat{\rho}_{S,I}(t')\otimes\hat{\rho}_{R,I}(0)\right]\right] \right) dt'.
\end{eqnarray} 
Importantly, this approximation does not imply the absence of excitations in the reservoir. Instead, it assumes that excitations generated by the interaction do not give rise to system-reservoir correlations strong enough to produce back-actions on the system's dynamics~\cite{Petruccione}. Within the Born approximation, the integro-differential equation of the system's reduced density operator retains only second-order contributions in the interaction Hamiltonian, which effectively leads to Eq.~\eqref{eq:1.21}. The validity of the approximation requires that the strengths of the system-reservoir couplings remain small compared to the intrinsic frequencies of the system: $g_{\mathbf{k},\lambda}^{(l)} \ll \omega_{lc}$.\\
\item {\bf Markov approximation:}\\
The \emph{Markov approximation} assumes that the decay time $\tau_R$ of the reservoir's autocorrelation function is much shorter than the characteristic relaxation time $\tau_S$ of the system, i.e., $\tau_R \ll \tau_S$~\cite{Petruccione}. Under this condition, memory effects are negligible: information transferred from the system to the reservoir is rapidly dissipated and does not flow back. 
Consequently, the time evolution of the system's reduced density operator becomes local in time, such that $\hat{\rho}_{S}(t)$ depends only on $\hat{\rho}_{S}(t)$ and not on its past history. This approximation leads to a so-called \emph{memoryless} or \emph{Markovian process}.
Mathematically, the contribution to the integral in Eq.~\eqref{eq:1.21} from times $t'< t-\tau_R$ is negligible, given that memory effects are no longer relevant for $t-t'>\tau_R$. One may therefore extend the upper limit of the integral in Eq.~\eqref{eq:1.21} to infinity and replace $\hat{\rho}_{S,I}(t')$ with $\hat{\rho}_{S,I}(t)$:
\begin{eqnarray}\label{eq:1.22}
\frac{d\hat{\rho}_{S,I}(t)}{dt} &\approx & -\frac{i}{\hbar}{\rm Tr}_R\left(\left[ \hat{H}_I(t), \hat{\rho}_{S,I}(0) \otimes \hat{\rho}_{R,I}(0) \right] \right) \nonumber \\
&-&\frac{1}{\hbar^2}\int_0^\infty {\rm Tr}_R\left( \left[ \hat{H}_I(t), \left[ \hat{H}_I(t'), \hat{\rho}_{S,I}(t)\otimes\hat{\rho}_{R,I}(0)\right]\right] \right) dt'.
\end{eqnarray}

The decay time $\tau_R$ of a radiation field is inversely proportional to its spectral bandwidth, $\tau_R \sim 1/\Delta\nu$, except special cases given by structured or multi-peaked spectra~\cite{Mandel}. Therefore, a broadband radiation field exhibits an extremely short correlation time. This has been quantitatively confirmed from studying thermal and solar radiations, whose optical bandwidth leads to correlation times on the order of a few femtoseconds~\cite{Ricketti}. 
\end{itemize}

Eq.~\eqref{eq:1.22}, where both the Born and Markov approximations are employed, is a \emph{Markovian quantum master equation} that is commonly known as \emph{Redfield} or \emph{Bloch-Redfield equation}~\cite{Petruccione}.

\subsubsection{Complete positivity of the Bloch-Redfield dynamics}\label{subsubsec:2.2}

The Bloch-Redfield (BR) equation \eqref{eq:1.22} provides a microscopic description of weak system-reservoir interactions, but complete positivity of the reduced density operator $\hat{\rho}_{S,I}(t)$ is not generally ensured~\cite{Dümcke, Kohen, Petruccione, Manzano, Gardiner}.
The lack of positivity can lead to unphysical predictions, such as negative populations. To restore positivity of $\hat{\rho}_{S,I}(t)$, it is customary to introduce additional approximations; typically, the \emph{secular approximation}~\cite{Petruccione,Manzano,McCauley,Pradilla}. The secular approximation removes rapidly time-oscillating terms in the interaction picture by performing a temporal coarse-grained average, which yields a generator in Lindblad form. However, secularization is not always well-justified or accurate~\cite{Pradilla,Jeske,Eastham}. In situations where the system's energy levels are nearly degenerate in relation to the system-environment coupling strength, the secular approximation neglects important quantum interference effects between these energy levels. As a result, the approximation suppresses quantum interference terms and fails to capture key dynamical features~\cite{Dodin-Brumer, McCauley, Pradilla}. These limitations are particularly severe in multi-level systems driven by incoherent or broadband radiation~\cite{Agarwal,Kozlov,Svidzinsky,Tscherbul-Brumer2,Tscherbul-Brumer,Dodin-Tscherbul,Dodin-Tscherbul2,Koyu-Tscherbul}, or in non-equilibrium conditions~\cite{Koyu-Dodin,Li,Ivander} where long-lived coherence effects can arise.

Several recent analyses have established that, for nearly degenerate optical transitions, the Bloch-Redfield equation obtained without imposing a full secular approximation remains physically reliable, whenever the {\it reservoir correlation matrix is positive definite} and the {\it non-secular couplings are supported by the dipole geometry}. In particular, Farina \emph{et al.}~\cite{Farina} show that the non-secular terms, inducing Fano coherence, do not compromise positivity when the radiation field is spectrally smooth on the scale of the level splitting $\Delta$. This results has been confirmed independently in Ref.~\cite{McCauley}.
Moreover, when $\Delta$ is small compared to the optical frequencies in input, the frequencies $\omega_{ac}$ and $\omega_{bc}$ of the system transitions are naturally comparable to, or smaller than, the inverse timescale of the reservoir autocorrelation function. As argued by Trushechkin~\cite{Trushechkin}, such transitions fall within the regime of applicability of a \emph{Unified Quantum Master Equation (UQME)} that retains a cluster of non-secular couplings while secularizing the coupling of all the other energy transitions. This construction yields a dynamical generator in Gorini-Kossakowski-Lindblad-Sudarshan (GKLS) form that preserves complete positivity and ensures thermodynamic consistency. Moreover, Gerry \emph{et al.}~\cite{Gerry} show that the UQME also satisfies quantum fluctuation relations for the heat transfer, thus confirming the physical soundness of the formalism.
Finally, the analysis of V-type systems subjected to a thermal environment presented in Ref.~\cite{Tscherbul} demonstrates that, for multi-level systems with closely-spaced energy transitions, retaining the appropriate non-secular terms leads to a Kossakowski matrix with non-negative eigenvalues, and thus a GKLS generator that can be written in a positive dissipative form. Taken all together, these results confirm that the Bloch-Redfield equation that retains interference terms associated with nearly degenerate excited states, preserves complete positivity and provides a physical description for quantum coherence generation thanks to incoherent input radiation in V-type systems.

%%%%%%%%%%%%%%%%%%%%%%%%%%%%%%%%
\subsubsection{Application of the Weisskopf-Wigner approximation}\label{subsubsec:3}

Let us proceed with further simplification of the Bloch-Redfield equation \eqref{eq:1.22}. The first term on the right-hand side of Eq.~\eqref{eq:1.22} is associated with the \emph{coherent part} of the system dynamics, which is driven by the interaction Hamiltonian $\hat{H}_I(t)$. We have denoted this term as $\frac{d\hat{\rho}_{S,I}(t)}{dt}|_{\text{coh}}$. By inserting $\hat{H}_I(t)$ in \eqref{eq:1.22}, it becomes:
\begin{equation}\label{eq:1.23}
\begin{aligned}
\left.\frac{d\hat{\rho}_{S,I}(t)}{dt}\right|_{\text{coh}} =& -\frac{i}{\hbar}{\rm Tr}_R\left(\left[ \hat{H}_I(t), \hat{\rho}_{S,I}(0) \otimes \hat{\rho}_{R,I}(0) \right]\right)\\
= & -\sum_{\lambda=1,2}\sum_{\bf{k}}g_{\bf{k},\lambda}^ae^{i(\omega_{ac}-\nu_{\bf{k},\lambda})t}\langle\hat{a}_{{\bf k},\lambda}\rangle[\hat{\sigma}_{ac}^+,\hat{\rho}_{S,I}(0)] \\
& - \sum_{\lambda=1,2}\sum_{\bf{k}} g_{\bf{k},\lambda}^be^{i(\omega_{bc}-\nu_{\bf{k},\lambda})t}\langle\hat{a}_{{\bf k},\lambda}\rangle[\hat{\sigma}_{bc}^+,\hat{\rho}_{S,I}(0)] + {\rm h.c.},
\end{aligned}
\end{equation}
where $\langle\hat{a}_{{\bf k},\lambda}\rangle \equiv {\rm Tr}_R\left(\hat{a}_{{\bf k},\lambda}\hat{\rho}_{R,I}(0)\right)$ are the expectation values of $\hat{a}_{{\bf k},\lambda}$ with respect to the reservoir's state $\hat{\rho}_{R,I}(0)$.

The thermal reservoir $R$ is in equilibrium at temperature $T$, which implies that its modes are distributed as a mixture of uncorrelated thermal equilibrium states at temperature $T$. Hence, the reservoir's state can be represented by the following reduced density operator~\cite{Scully:Zub,Petruccione}:
\begin{equation}\label{eq:1.24}
\hat{\rho}_R(0) = \prod_{\bf{k},\lambda}\left[1-\exp\left(-\frac{\hbar\nu_{\bf{k},\lambda}}{k_BT}\right)\right]\exp\left(-\frac{\hbar\nu_{\bf{k},\lambda}\hat{a}_{{\bf k},\lambda}^\dagger\hat{a}_{{\bf k},\lambda}}{k_BT}\right), 
\end{equation}
where $k_B$ is the Boltzmann constant.
In this way, the expectation value and the correlation function of the reservoir's operators, computed with respect to $\hat{\rho}_R(0)$, take the following expressions: 
\begin{eqnarray}
\langle\hat{a}_{{\bf k},\lambda}\rangle &=&\langle\hat{a}_{{\bf k},\lambda}^\dagger\rangle = 0 \quad \forall \, {\bf k},\lambda \label{eq:1.25a}\\ 
\langle\hat{a}_{{\bf k},\lambda}^\dagger\hat{a}_{{\bf k'},\lambda'}\rangle &=& \bar{n}_{{\bf k},\lambda}\delta_{{\bf k},{\bf k'}}\delta_{\lambda,\lambda'}\label{eq:1.25b}\\
\langle\hat{a}_{{\bf k},\lambda}\hat{a}_{{\bf k'},\lambda'}^\dagger\rangle &=& (\bar{n}_{{\bf k},\lambda}+1)\delta_{{\bf k},{\bf k'}}\delta_{\lambda,\lambda'}\label{eq:1.25c}\\
\langle\hat{a}_{{\bf k},\lambda}\hat{a}_{{\bf k'},\lambda'}\rangle &=& \langle\hat{a}_{{\bf k},\lambda}^\dagger\hat{a}_{{\bf k'},\lambda'}^\dagger\rangle=0 \quad \forall \, {\bf k,k'},\lambda,\lambda'\, , \label{eq:1.25d}
\end{eqnarray}
where 
\begin{equation}\label{eq:1.25e}
\bar{n}_{{\bf k},\lambda} = \frac{1}{\exp\left(\frac{\hbar\nu_{{\bf k},\lambda}}{k_B T}\right)-1}
\end{equation}
is the average photon number in the mode ${\bf k}$ with polarization $\lambda$.
These relations indicate that in thermal equilibrium, the expectation values of the annihilation and creation operators are equal to zero (Eq.~\eqref{eq:1.25a}). Moreover, Eqs.~\eqref{eq:1.25b}-\eqref{eq:1.25c} give the average occupation numbers, while Eq.~\eqref{eq:1.25d} shows that there are no correlations between the modes. Eq.~\eqref{eq:1.25a} implies that the coherent part of the reduced dynamics of the system is equal to zero:
\begin{equation}\label{eq:1.26} \left.\frac{d\hat{\rho}_{S,I}(t)}{dt}\right|_{\text{coh}}=0.
\end{equation}

Let us now consider the second term on the right-hand side of Eq.~\eqref{eq:1.22}, which is associated with the \emph{incoherent part} of the reduced dynamics:
\begin{equation}\label{eq:1.27}
\left.\frac{d\hat{\rho}_{S,I}(t)}{dt}\right|_{\text{incoh}}=-\frac{1}{\hbar^2}\int_0^\infty {\rm Tr}_R \left(\left[ \hat{H}_I(t), \left[ \hat{H}_I(t'), \hat{\rho}_{S,I}(t)\otimes\hat{\rho}_{R,I}(0)\right]\right] \right) dt'.
\end{equation}
This term is related to the interaction between the system and reservoir, which leads to dissipation. Eq.~\eqref{eq:1.27} can be simplified by expanding the double commutator after substituting Eq.~\eqref{eq:1.16} in Eq.~\eqref{eq:1.27} and setting $t'=t-s$, where $s>\tau_R$ according to the Markov approximation. Each term in Eq.~\eqref{eq:1.27} is examined separately.\\
The first term is:
\begin{eqnarray}\label{eq:1.28a}
&&-\int_0^\infty \sum_{\lambda,\lambda'} \sum_{\bf{k},\bf{k'}} g_{\bf{k},\lambda}^{(l)}g_{\bf{k'},\lambda'}^{(l)}\,e^{-i(\omega_{lc}-\nu_{\bf{k},\lambda})t+i(\omega_{lc}-\nu_{\bf{k'},\lambda'})(t-s)}\Big[\langle\hat{a}_{\bf{k},\lambda}^\dagger\hat{a}_{\bf{k'},\lambda'}\rangle\Big( \hat{\sigma}_{lc}^{-}\hat{\sigma}_{lc}^{+}\hat{\rho}_{S,I}(t)\nonumber \\
&&- \sigma_{lc}^+\rho_{S,I}(t)\sigma_{lc}^-\Big) +\langle\hat{a}_{\bf{k'},\lambda'}\hat{a}_{\bf{k}\lambda}^\dagger\rangle\Big(\hat{\rho}_{S,I}(t)\hat{\sigma}_{lc}^{+}\hat{\sigma}_{lc}^{-} - \hat{\sigma}_{lc}^{-}\hat{\rho}_{S,I}(t)\hat{\sigma}_{lc}^+\Big)\Big] ds 
\end{eqnarray}
that is associated with the system's transition $|a\rangle\leftrightarrow|c\rangle$ with $l=a$ or $|b\rangle\leftrightarrow|c\rangle$ with $l=b$.\\
Also the following crossing terms, involving both the levels $|a\rangle$ and $|b\rangle$, arise in Eq.~\eqref{eq:1.27}: 
\begin{eqnarray}\label{eq:1.28b}
&&-\int_0^\infty \sum_{\lambda,\lambda'} \sum_{\bf{k},\bf{k'}} g_{\bf{k},\lambda}^{(a)}g_{\bf{k'},\lambda'}^{(b)}\,e^{i(\omega_{ac}-\nu_{\bf{k},\lambda})t-i(\omega_{bc}-\nu_{\bf{k'},\lambda'})(t-s)}\Big[\langle\hat{a}_{\bf{k'},\lambda'}^\dagger\hat{a}_{\bf{k},\lambda}\rangle\Big(\hat{\rho}_{S,I}(t)\hat{\sigma}_{bc}^-\hat{\sigma}_{ac}^{+} \nonumber\\
&&- \hat{\sigma}_{ac}^{+}\hat{\rho}_{S,I}(t)\hat{\sigma}_{bc}^-\Big)+
\langle\hat{a}_{\bf{k},\lambda}\hat{a}_{\bf{k'},\lambda'}^\dagger\rangle\Big(\hat{\sigma}_{ac}^+\hat{\sigma}_{bc}^-\hat{\rho}_{S,I}(t) - \hat{\sigma}_{bc}^-\hat{\rho}_{S,I}(t)\hat{\sigma}_{ac}^+\Big)\Big] ds \,.
\end{eqnarray}
Hence, from substituting the expectation values in Eqs.~\eqref{eq:1.25b}-\eqref{eq:1.25d}, Eqs.~\eqref{eq:1.28a}-\eqref{eq:1.28b} simplify as
\begin{eqnarray}\label{eq:1.29a}
&&-\int_0^\infty \sum_{\lambda=1,2} \sum_{\bf{k}} |g_{\bf{k},\lambda}^{(l)}|^2e^{-i(\omega_{lc}-\nu_{\bf{k},\lambda})s}\Big[\bar{n}_{\bf{k},\lambda}\left(\hat{\sigma}_{lc}^-\hat{\sigma}_{lc}^+\hat{\rho}_{S,I}(t)-\hat{\sigma}_{lc}^+\hat{\rho}_{S,I}(t)\hat{\sigma}_{lc}^-\right)\nonumber \\
&& + (\bar{n}_{\bf{k},\lambda}+1)\left( \hat{\rho}_{S,I}(t)\hat{\sigma}_{lc}^+\hat{\sigma}_{lc}^{-} - \hat{\sigma}_{lc}^{-}\hat{\rho}_{S,I}(t)\hat{\sigma}_{lc}^+ \right)\Big] ds 
\end{eqnarray}
and
\begin{eqnarray}\label{eq:1.29b}
&&-\int_0^\infty \sum_{\lambda=1,2} \sum_{\bf{k}} g_{\bf{k},\lambda}^{(a)}g_{\bf{k},\lambda}^{(b)}\,e^{i(\omega_{ac}-\nu_{\bf{k},\lambda})t-i(\omega_{bc}-\nu_{\bf{k},\lambda})(t-s)}\Big[\bar{n}_{\bf{k},\lambda}\left(\hat{\rho}_{S,I}(t)\hat{\sigma}_{bc}^-\hat{\sigma}_{ac}^+ - \hat{\sigma}_{ac}^+\hat{\rho}_{S,I}(t)\hat{\sigma}_{bc}^-\right)\nonumber \\
&& + (\bar{n}_{\bf{k},\lambda}+1)\left(\hat{\sigma}_{ac}^+\hat{\sigma}_{bc}^-\hat{\rho}_{S,I}(t) - \hat{\sigma}_{bc}^-\hat{\rho}_{S,I}(t)\hat{\sigma}_{ac}^+\right)\Big] ds = \nonumber \\
&& = - \int_0^\infty \sum_{\lambda=1,2} \sum_{\bf{k}} g_{\bf{k},\lambda}^{(a)}g_{\bf{k},\lambda}^{(b)}\,e^{i\Delta t }e^{i(\omega_{bc}-\nu_{\bf{k},\lambda})s}\Big[\bar{n}_{\bf{k},\lambda}\left(\hat{\rho}_{S,I}(t)\hat{\sigma}_{bc}^-\hat{\sigma}_{ac}^+ - \hat{\sigma}_{ac}^+\hat{\rho}_{S,I}(t)\hat{\sigma}_{bc}^-\right)\nonumber \\
&& + (\bar{n}_{\bf{k},\lambda}+1)\left(\hat{\sigma}_{ac}^+\hat{\sigma}_{bc}^-\hat{\rho}_{S,I}(t) - \hat{\sigma}_{bc}^-\hat{\rho}_{S,I}(t)\hat{\sigma}_{ac}^+\right)\Big] ds\,,
\end{eqnarray}
where we have made use of $\omega_{ac} -\omega_{bc} = \Delta$. At this point, the \emph{Weisskopf-Wigner approximation} can be applied.
\begin{itemize}
\item {\bf Weisskopf-Wigner approximation:}\\
The Weisskopf-Wigner approximation assumes that the electromagnetic field's modes are densely distributed in the frequency domain. This assumption allows to replace the discrete sum over the wavevectors ${\bf k}$ by an integral over a continuous density of modes in the ${\bf k}$-space~\cite{Scully:Zub,Petruccione}:
\begin{equation}\label{eq:1.30a}
    \sum_{\bf{k}}\longrightarrow \frac{V}{(2\pi)^3} \int_0^\infty d^3{\bf k}. 
\end{equation}

In the Weisskopf-Wigner framework, the spontaneous emission is customarily described by the coupling of the system with the electromagnetic vacuum, which is isotropic in both the propagation direction and the transverse polarization. Such an isotropy means angularly-uniform emission patterns, which justifies performing the continuum limit over the full three-dimensional ${\bf k}$-space. When the incoherent radiation in input is such that, absorption, stimulated and spontaneous emissions occur with the same isotropic mode density, then all these processes can be effectively described in terms of the interaction with a single electromagnetic reservoir. As a result, the integral in Eq.~\eqref{eq:1.30a} can be calculated by representing ${\bf k}$ in spherical coordinates: ${\bf k} = |{\bf k}|(\sin\theta\cos\phi,\sin\theta\sin\phi,\cos\theta)$, which varies over the infinitesimal spherical volume $d^3{\bf k}=|{\bf k}|^2d|{\bf k}|\sin\theta d\theta d\phi$. In this way, replacing $|{\bf k}|$ with $\frac{ \nu_{\bf k}}{c}$ (dispersion relation), one has:
\begin{equation}\label{eq:1.30b}
    \sum_{\bf{k}}\longrightarrow \frac{V}{(2\pi c)^3} \int_0^{2\pi}d\phi\int_{0}^\pi d\theta\sin\theta\int_0^\infty \nu_{\bf k}^2d\nu_{\bf k}. 
\end{equation}
Using Eq. \eqref{eq:1.30b}, the spectral density, defined in Eq. \eqref{eq:1.17}, becomes: 
\begin{eqnarray}\label{eq:1.31a}
      J_l(\nu)&\approx &\frac{V}{(2\pi c)^3} \sum_{\lambda=1,2} \int_0^{2\pi}d\phi\int_{0}^\pi d\theta\sin\theta\int_0^\infty |g_{\bf{k},\lambda}^{(l)}|^2\delta(\nu-n) n^{2}dn \nonumber\\
      & = & \frac{\nu^3}{16\hbar\pi^3\varepsilon_oc^3} \sum_{\lambda=1,2} \int_0^{2\pi}d\phi\int_{0}^\pi |\bm{\mu}_{lc}\cdot\bm{\epsilon}_{\bf{k},\lambda}|^2 \sin\theta d\theta, 
\end{eqnarray}  
where we have used both the properties of the Dirac delta function and the definition of Eq.~\eqref{eq:1.15}.

In the presence of broadband incoherent radiation, the spectral density can be treated as locally flat around the system transitions, provided that its bandwidth exceeds both the excited state splitting and the inverse relaxation time.
It allows us to set, for any $\lambda$, the conditions $\nu_{\bf{k},\lambda} \approx \omega_{ac}$ and $\approx \omega_{bc}$ for all $\bf{k}$ that are compatible with the energies of the transitions $|a\rangle \leftrightarrow |c\rangle$ and $|b\rangle \leftrightarrow |c\rangle$, respectively. Consequently, as the term $\nu^3$ is slowly varying near $\omega_{lc}$, it can be substituted by $\omega_{lc}^3$~\cite{Loudon,Scully:Zub} in Eq.~\eqref{eq:1.31a}, such that
\begin{equation}\label{eq:1.31b}
     J_l(\omega_{lc})=\frac{\omega_{lc}^3}{16\hbar\pi^3\varepsilon_oc^3} \sum_{\lambda=1,2} \int_0^{2\pi}d\phi\int_{0}^\pi |\bm{\mu}_{lc}\cdot\bm{\epsilon}_{\bf{k},\lambda}|^{2} \sin\theta d\theta.
\end{equation} 

Under the Weisskopf-Wigner approximation and the condition $\omega_{ac},\omega_{bc}\gg\Delta$, the thermal occupation number can be evaluated at the mean frequency
$\frac{\omega_{ac}+\omega_{bc}}{2}=\omega_{bc} + \frac{\Delta}{2}=\omega_{ac} - \frac{\Delta}{2}$ for any mode ${\bf k}$, i.e.,
\begin{equation}\label{eq:1.38}
\bar{n} = \frac{1}{\exp\left(\frac{\hbar(\omega_{ac}+\omega_{bc})}{2k_B T}\right)-1}\,.
\end{equation}
Also the polarization vector can be expressed in spherical coordinates in order to compute the angular integrals in Eq.~\eqref{eq:1.31b}~\cite{Koyu-Dodin,Loudon}. Following the methodology used in~\cite{Koyu-Dodin,Loudon,Donati}, the following two instances ${\bm\epsilon}_{1}$, ${\bm\epsilon}_{2}$ of the polarization vector for $\lambda=1,2$ hold: 
\begin{eqnarray}
&& {\bm\epsilon}_{1} =[-\cos\theta\cos\phi,-\cos\theta\sin\phi,\sin\theta]\label{eq:1.31c} \\
&& {\bm\epsilon}_{2}=[\sin\phi,-\cos\phi,0],\label{eq:1.31d} 
\end{eqnarray}
which satisfy the transversality condition of Eq.~\eqref{eq:1.6} and the orthonormality relation of Eq.~\eqref{eq:1.7}.
Therefore, by calculating the scalar products $\left(\bm{\mu}_{lc}\cdot\bm{\epsilon}_{1}\right)$ and $\left(\bm{\mu}_{lc}\cdot\bm{\epsilon}_{2}\right)$ ($l=a,b$) for arbitrary electric dipole moments $\bm{\mu}_{ac}$ and $\bm{\mu}_{bc}$, and by evaluating the integrals over the spherical polar angles $\theta$ and $\phi$, Eq.~\eqref{eq:1.31b} simplifies to
\begin{eqnarray}\label{eq:1.31e}
J_l(\omega_{lc}) &=& \frac{\omega_{lc}^3}{16\hbar\pi^3\varepsilon_{o} c^3} \int_0^{2\pi}d\phi\int_0^\pi\sin\theta d\theta\sum_{\lambda=1,2} \left|\bm{\mu}_{lc}\cdot\bm{\epsilon}_{\bf{k},\lambda}\right|^{2}\nonumber \\
&=& \frac{\omega_{lc}^3}{16\hbar\pi^3\varepsilon_{o} c^3}\frac{8}{3}\pi \left|\bm{\mu}_{lc}\right|^2 
= \frac{\omega_{lc}^3}{6\hbar\pi^2\varepsilon_{o} c^3}\left|\bm{\mu}_{lc}\right|^{2}.
\end{eqnarray}
Moreover, the cross-term $\sqrt{J_a(\omega_{ac})J_b(\omega_{bc})}$ becomes: 
\begin{eqnarray}\label{eq:1.31f}
\sqrt{J_a(\omega_{ac})J_b(\omega_{bc})} &=& \frac{\omega_{ac}^3}{16\hbar\pi^3\varepsilon_oc^3} \int_0^{2\pi}d\phi\int_0^\pi\sin\theta d\theta\sum_{\lambda=1,2} \left(\bm{\mu}_{ac}\cdot\bm{\epsilon}_{\bf{k},\lambda}\right)\left(\bm{\mu}_{bc}\cdot\bm{\epsilon}_{\bf{k},\lambda}\right)\nonumber \\
&=&\frac{\omega_{ac}^3}{16\hbar\pi^3\varepsilon_oc^3}\frac{8}{3}\pi\left(\bm{\mu}_{ac}\cdot\bm{\mu}_{bc}\right)
= \frac{\omega_{ac}^3}{6\hbar\pi^2\varepsilon_oc^3}\pi\left(\bm{\mu}_{ac}\cdot\bm{\mu}_{bc}\right).
\end{eqnarray}

By contrast, when the external radiation field is restricted to a subset of propagation directions (e.g., an anisotropic radiation), the modes of the reservoir must be treated explicitly in the frequency domain~\cite{Dodin-Brumer,Koyu-Dodin}. This involves an effective separation between modes that lead to isotropic spontaneous emission and modes that lead to anisotropic absorption and stimulated emission, respectively. Isotropic spontaneous emission can be modelled by the interaction of the system with the electromagnetic vacuum, whereas anisotropic absorption and stimulated emission are given by the interaction with a directional excitation thermal reservoir. As a result, in the latter, the integration over the spherical volume element $d^3{\bf k}=|{\bf k}|^2 d|{\bf k}| \sin\theta\, d\theta\, d\phi$ reduces to an integration over the magnitude of the wavevector only, which yields:
\begin{equation}\label{eq:1.31g}
    \sum_{\bf{k}}\longrightarrow \frac{V}{(2\pi)^3} \int_0^\infty |{\bf k}|^2\, d|{\bf k}|.
\end{equation}
Accordingly, the spectral density associated with the directional excitation thermal reservoir is $J_l(\nu)=\frac{\nu^3}{16\hbar\pi^3\varepsilon_0 c^3} \sum_{\lambda=1,2} \left|\bm{\mu}_{lc}\cdot\bm{\epsilon}_{\bf{k},\lambda}\right|^{2}$, which can be further approximated as
\begin{equation}\label{eq:1.31i}
  J_l(\omega_{lc})=\frac{\omega_{lc}^3}{16\hbar\pi^3\varepsilon_0 c^3}
  \sum_{\lambda=1,2} \left|\bm{\mu}_{lc}\cdot\bm{\epsilon}_{\bf{k},\lambda}\right|^{2}.
\end{equation}
The validity of Eq.~\eqref{eq:1.31i} is justified by the fact that the spectral bandwidth of the anisotropic radiation is flat in the vicinity of atomic transitions under the Weisskopf-Wigner approximation. Note that the corresponding average photon occupation number $\bar{n}$ is given by Eq.~\eqref{eq:1.38}, as in the case of isotropic input radiations. This is because $\bar{n}$ depends only on the spectral properties of the input radiation and not on the angular distribution of its modes' wavevectors in the spatial domain.
\end{itemize}

The result of applying the Weisskopf-Wigner approximation in Eqs.~\eqref{eq:1.29a}-\eqref{eq:1.29b} is to write them as
\begin{eqnarray}\label{eq:1.32a}
    && - \int_0^\infty  \int_0^\infty J_l(\omega_{lc}) e^{-i(\omega_{lc}-\nu_{\bf{k},\lambda})s}\Big[\bar{n}\left(\hat{\sigma}_{lc}^-\hat{\sigma}_{lc}^+\hat{\rho}_{S,I}(t)-\hat{\sigma}_{lc}^+\hat{\rho}_{S,I}(t)\hat{\sigma}_{lc}^-\right)\nonumber\\
    && +(\bar{n}+1)\left( \hat{\rho}_{S,I}(t)\hat{\sigma}_{lc}^+\hat{\sigma}_{lc}^{-} - \hat{\sigma}_{lc}^{-}\hat{\rho}_{S,I}(t)\hat{\sigma}_{lc}^+ \right)\Big] d\nu_{\bf{k},\lambda}ds\nonumber \\
    &&
\end{eqnarray}
and
\begin{eqnarray}\label{eq:1.32b}
    && - \int_0^\infty  \int_0^\infty \sqrt{J_a(\omega_{ac})J_b(\omega_{bc})} e^{i\Delta t }e^{i(\omega_{bc}-\nu_{\bf{k},\lambda})s}\Big[\bar{n}\left(\hat{\rho}_{S,I}(t)\hat{\sigma}_{bc}^-\hat{\sigma}_{ac}^+ - \hat{\sigma}_{ac}^+\hat{\rho}_{S,I}(t)\hat{\sigma}_{bc}^-\right)\nonumber \\
    && + (\bar{n}+1)\left(\hat{\sigma}_{ac}^+\hat{\sigma}_{bc}^-\hat{\rho}_{S,I}(t) - \hat{\sigma}_{bc}^-\hat{\rho}_{S,I}(t)\hat{\sigma}_{ac}^+\right)\Big] d\nu_{\bf{k},\lambda}ds\,.
\end{eqnarray}      
Notice that when $\nu_{\bf{k},\lambda}\neq\omega_{lc}$, the exponential terms $e^{i(\omega_{lc}-\nu_{\bf{k},\lambda})s}$ in Eqs.~\eqref{eq:1.32a}-\eqref{eq:1.32b} oscillate rapidly. According to the Weisskopf-Wigner approximation, the only relevant frequency modes are those close to the system's transitions. Eqs.~\eqref{eq:1.32a}-\eqref{eq:1.32b} can be further simplified after have inserted the following integral computations:
\begin{equation}\label{eq:1.33a}
    \int_0^\infty e^{-i(\omega_{lc}-\nu_{\bf{k},\lambda})s} d\nu_{\bf{k},\lambda} = e^{-i\omega_{lc}s}\int_0^\infty e^{i\nu_{\bf{k},\lambda}s}d\nu_{\bf{k},\lambda} = e^{-i\omega_{lc}s}\left( \pi\delta (s) +i\mathbb{P}\frac{1}{s}\right) 
\end{equation}
and 
\begin{equation}\label{eq:1.33b}
    \int_0^\infty e^{i\Delta t }e^{i(\omega_{bc}-\nu_{\bf{k},\lambda})s} d\nu_{\bf{k},\lambda}= e^{i\Delta t }e^{i\omega_{bc}s}\int_0^\infty e^{-i\nu_{\bf{k},\lambda}s} d\nu_{\bf{k},\lambda}
    = e^{i\Delta t }e^{i\omega_{bc}s}\left( \pi\delta (s) -i\mathbb{P}\frac{1}{s}\right),
\end{equation}
where we have exploited the one-side Fourier transform of the Dirac delta function, i.e.,
\begin{equation}\label{eq:1.34}
\int_0^\infty e^{\pm i\nu s}d\nu\equiv\pi\delta (s) \pm i\mathbb{P}\frac{1}{s}.
\end{equation} 
The term $\mathbb{P}$ denotes the Cauchy principal value of the integral, which regularizes the divergence of the integral due to the singularity at $s=0$. This term gives rise to the \emph{Lamb shift effect}, which origins from the interaction of the system with the vacuum fluctuations of the electromagnetic field. This effect causes a slight shift in the energy levels of the system~\cite{Petruccione,McCauley}. In our analysis, the Lamb shift term can be neglected, as also discussed in Refs.~\cite{Tscherbul-Brumer,Dodin-Tscherbul2,Koyu-Dodin}, since it is  negligible for weak system-radiation couplings, being much smaller than the transition frequencies of the system.

Substituting expressions \eqref{eq:1.33a} and \eqref{eq:1.33b} in Eqs.~\eqref{eq:1.32a}-\eqref{eq:1.32b} yields:
\begin{equation}\label{eq:1.35a}
- \pi J_l(\omega_{lc}) \int_0^\infty e^{-i\omega_{lc}s}\Big[\bar{n}\left(\hat{\sigma}_{lc}^-\hat{\sigma}_{lc}^+\hat{\rho}_{S,I}(t)-\hat{\sigma}_{lc}^+\hat{\rho}_{S,I}(t)\hat{\sigma}_{lc}^-\right)
+(\bar{n}+1)\left( \hat{\rho}_{S,I}(t)\hat{\sigma}_{lc}^+\hat{\sigma}_{lc}^{-} - \hat{\sigma}_{lc}^{-}\hat{\rho}_{S,I}(t)\hat{\sigma}_{lc}^+ \right)\Big]\delta (s) ds 
\end{equation}
with $l=a,b$, and
\begin{eqnarray}\label{eq:1.35b}
&& - \pi \sqrt{J_a(\omega_{ac})J_b(\omega_{bc})}\int_0^\infty  e^{i\Delta t }e^{i\omega_{bc}s}\Big[\bar{n}\left(\hat{\rho}_{S,I}(t)\hat{\sigma}_{bc}^-\hat{\sigma}_{ac}^+ - \hat{\sigma}_{ac}^+\hat{\rho}_{S,I}(t)\hat{\sigma}_{bc}^-\right)\nonumber \\
&& + (\bar{n}+1)\left(\hat{\sigma}_{ac}^+\hat{\sigma}_{bc}^-\hat{\rho}_{S,I}(t) - \hat{\sigma}_{bc}^-\hat{\rho}_{S,I}(t)\hat{\sigma}_{ac}^+\right)\Big] \delta (s) ds\,.
\end{eqnarray} 
By evaluating the integral, these terms are equal to
\begin{equation}\label{eq:1.36a}
- \pi J(\omega_{lc})\Big[\bar{n}\left(\hat{\sigma}_{lc}^-\hat{\sigma}_{lc}^+\hat{\rho}_{S,I}(t)-\hat{\sigma}_{lc}^+\hat{\rho}_{S,I}(t)\hat{\sigma}_{lc}^-\right)
+(\bar{n}+1)\left( \hat{\rho}_{S,I}(t)\hat{\sigma}_{lc}^+\hat{\sigma}_{lc}^{-} - \hat{\sigma}_{lc}^{-}\hat{\rho}_{S,I}(t)\hat{\sigma}_{lc}^+ \right)\Big] 
\end{equation}
and
\begin{equation}\label{eq:1.36b}
- \pi \sqrt{J(\omega_{ac})J(\omega_{bc})}  e^{i\Delta t }\Big[\bar{n}\left(\hat{\rho}_{S,I}(t)\hat{\sigma}_{bc}^-\hat{\sigma}_{ac}^+ - \hat{\sigma}_{ac}^+\hat{\rho}_{S,I}(t)\hat{\sigma}_{bc}^-\right)
+ (\bar{n}+1)\left(\hat{\sigma}_{ac}^+\hat{\sigma}_{bc}^-\hat{\rho}_{S,I}(t) - \hat{\sigma}_{bc}^-\hat{\rho}_{S,I}(t)\hat{\sigma}_{ac}^+\right)\Big].
\end{equation} 

\subsubsection{Application of the partial secular approximation}

Eq.~\eqref{eq:1.36b} includes an oscillatory component at the frequency $\Delta$. If the two excited states of the system are nearly degenerate, then the oscillation period  $1/\Delta$ can exceed the system's characteristic timescale. In such a case, the oscillatory component at frequency $\Delta$ can not be averaged out, so as not to risk neglecting relevant interference effects entering the dynamics of the system. The application of the \emph{partial secular approximation} ensures that this condition is met. 
\begin{itemize}
\item {\bf Partial secular approximation:}\\
The partial secular approximation retains all terms oscillating at frequencies of order $\Delta$, while averaging out those oscillating at optical frequencies $\omega_{ac}$ and $\omega_{bc}$. This approach, widely used in the study of coherence generation by incoherent radiation~\cite{Tscherbul-Brumer,Dodin-Brumer,Li,Jeske,Eastham}, preserves the interference between the two radiative pathways, from excited to ground states and vice versa.

The justification of this application relies on the near-degeneracy condition $\Delta \ll \omega_{ac},\omega_{bc}$, which for example is well satisfied for optical transitions where $\omega_{ac},\omega_{bc}\sim 100\,\mathrm{THz}$. Under the partial secular approximation, the rapidly oscillating terms proportional to $e^{\pm i\omega_{ac} t}$ and $e^{\pm i\omega_{bc} t}$ average out on the system's timescale, while terms oscillating at $e^{\pm i\Delta t}$ evolve slower and must be retained. Moreover, as $\Delta \ll \omega_{ac},\omega_{bc}$, we can set $e^{i\Delta t} \approx 1$ for any $t$.
\end{itemize}
After the application of the partial secular approximation, the crossing term \eqref{eq:1.36b} becomes:
\begin{equation}\label{eq:1.37a}
- \pi \sqrt{J_a(\omega_{ac})J_b(\omega_{ac})}  \Big[\bar{n}\left(\hat{\rho}_{S,I}(t)\hat{\sigma}_{bc}^-\hat{\sigma}_{ac}^+ - \hat{\sigma}_{ac}^+\hat{\rho}_{S,I}(t)\hat{\sigma}_{bc}^-\right)+(\bar{n}+1)\left(\hat{\sigma}_{ac}^+\hat{\sigma}_{bc}^-\hat{\rho}_{S,I}(t) - \hat{\sigma}_{bc}^-\hat{\rho}_{S,I}(t)\hat{\sigma}_{ac}^+\right)\Big].
\end{equation} 

\subsubsection{Isotropic incoherent radiation field}

Let us now provide the expression of the master equation for a quantum system interacting with an isotropic incoherent radiation. For this purpose, we substitute Eqs.~\eqref{eq:1.36a}-\eqref{eq:1.37a} into the Born-Redfield equation \eqref{eq:1.22}, so that:
{\small
\begin{eqnarray}\label{eq:1.39}
&&\frac{d\hat{\rho}_{S,I}(t)}{dt} = \left.\frac{d\hat{\rho}_{S,I}(t)}{dt}\right|_{\text{incoh}} =\nonumber \\ 
&& - \pi J_a(\omega_{ac})\Big[ \bar{n}\big(\hat{\sigma}_{ac}^-\hat{\sigma}_{ac}^+\hat{\rho}_{S,I}(t)-\hat{\sigma}_{ac}^+\hat{\rho}_{S,I}(t)\hat{\sigma}_{ac}^-\big)+(\bar{n}+1)\big(\hat{\rho}_{S,I}(t)\hat{\sigma}_{ac}^+\hat{\sigma}_{ac}^{-} + 
 - \hat{\sigma}_{ac}^-\hat{\rho}_{S,I}(t)\hat{\sigma}_{ac}^+\big) \Big]\nonumber \\
&& - \pi J_b(\omega_{ac})\Big[ \bar{n}\big(\hat{\sigma}_{bc}^-\hat{\sigma}_{bc}^+\hat{\rho}_{S,I}(t)-\hat{\sigma}_{bc}^+\hat{\rho}_{S,I}(t)\hat{\sigma}_{bc}^-\big)+(\bar{n}+1)\big(\hat{\rho}_{S,I}(t)\hat{\sigma}_{bc}^+\hat{\sigma}_{bc}^- - \hat{\sigma}_{bc}^-\hat{\rho}_{S,I}(t)\hat{\sigma}_{bc}^+\big) \Big] \nonumber \\ 
&& - \pi \sqrt{J_a(\omega_{ac})J_b(\omega_{ac})} \bar{n}\big(\hat{\rho}_{S,I}(t)\hat{\sigma}_{bc}^-\hat{\sigma}_{ac}^+ - \hat{\sigma}_{ac}^+\hat{\rho}_{S,I}(t)\hat{\sigma}_{bc}^- \big) \nonumber \\
&& - \pi \sqrt{J_a(\omega_{ac})J_b(\omega_{ac})}
(\bar{n}+1)\big(\hat{\sigma}_{ac}^+\hat{\sigma}_{bc}^-\hat{\rho}_{S,I}(t) - \hat{\sigma}_{bc}^-\hat{\rho}_{S,I}(t)\hat{\sigma}_{ac}^+\big) \nonumber \\
&& - \pi \sqrt{J_a(\omega_{ac})J_b(\omega_{ac})} \bar{n}\big(\hat{\rho}_{S,I}(t)\hat{\sigma}_{ac}^-\hat{\sigma}_{bc}^+ - \hat{\sigma}_{bc}^+\hat{\rho}_{S,I}(t)\hat{\sigma}_{ac}^- \big) \nonumber \\
&& - \pi \sqrt{J_a(\omega_{ac})J_b(\omega_{ac})}(\bar{n}+1)\big(\hat{\sigma}_{bc}^+\hat{\sigma}_{ac}^-\hat{\rho}_{S,I}(t) - \hat{\sigma}_{ac}^-\hat{\rho}_{S,I}(t)\hat{\sigma}_{bc}^+\big) \Big] + \,\rm{h.c.},
\end{eqnarray}
}
where $J_l(\omega_{lc})$ ($l=a,b$) and $\sqrt{J_a(\omega_{ac})J_b(\omega_{ac})}$ are given by Eqs.~\eqref{eq:1.31e}-\eqref{eq:1.31f} respectively. The interaction with an isotropic (and unpolarized), incoherent field has been analysed also in Refs.~\cite{Kozlov,Tscherbul-Brumer2,Tscherbul-Brumer,Dodin-Tscherbul,Koyu-Tscherbul,Dodin-Tscherbul2,Donati}.

The master equation \eqref{eq:1.39} departs from the standard GKLS form due to the quantum coherence terms (the last four lines), which arise from the interference between the two optical transition pathways. As discussed in Subsection~\ref{subsubsec:2.2}, such non-secular terms do not necessarily compromise complete positivity when (i) they involve a cluster of nearly degenerate transitions, and (ii) the radiation spectrum is sufficiently flat on the scale of the excited state splitting~\cite{McCauley,Farina}. The quantum modelling reported in this paper includes the frequency-clustering regime analysed by Trushechkin~\cite{Trushechkin}. Further support for the complete positivity is provided by Jeske \emph{et al.}~\cite{Jeske}, demonstrating that, in this near-degenerate limit, the Bloch-Redfield generator reduces to a degenerate Lindblad generator, with the dissipator admitting a positive Kossakowski representation. For the specific case of a V-type quantum system, this conclusion is confirmed in Ref.~\cite{Tscherbul}.

Consistently with earlier analyses of noise-induced Fano coherence~\cite{Koyu-Tscherbul,Dodin-Brumer,Tscherbul-Brumer2}, we have also observed that the partial secular Bloch-Redfield equation \eqref{eq:1.39} preserves positivity throughout the parameter regimes relevant for quantum coherence generation in V-type atomic systems; see also Subsection~\ref{subsec:3.1} below. 

%%%%%%%%%%%%%%%%%%%%%%%%%%%%%%%%
\subsubsection{Anisotropic and polarized incoherent radiation field}\label{subsec:3}

Let us now consider an \emph{anisotropic} and \emph{polarized} incoherent input radiation field. Anisotropy influences the incoherent pumping terms, and therefore affects absorption and stimulated emission processes. 
Spontaneous emission remains isotropic, as it is primarily governed by the coupling of the system to vacuum electromagnetic fluctuations, which are isotropic in free space.
Following the approach introduced by Dodin \emph{et al.}~\cite{Dodin-Brumer} and already discussed in Subsec.~\ref{subsubsec:3}, this physical scenario can be modelled effectively by describing the system as interacting with two distinct thermal (photonic) reservoirs: (i) an isotropic vacuum reservoir, responsible for spontaneous decay processes, and (ii) a directional excitation thermal reservoir, which predominantly drives absorption and stimulated emissions. Using this representation, the dissipative terms in Eqs.~\eqref{eq:1.36a}-\eqref{eq:1.37a} are rewritten as the sum of isotropic and anisotropic contributions. For the diagonal terms, one obtains:
\begin{equation}\label{eq:1.40a}
- \pi J_{l,\rm iso}(\omega_{lc}) \Big[\bar{n}_{\rm iso}\left(\hat{\sigma}_{lc}^-\hat{\sigma}_{lc}^+\hat{\rho}_{S,I}(t)-\hat{\sigma}_{lc}^+\hat{\rho}_{S,I}(t)\hat{\sigma}_{lc}^-\right) + (\bar{n}_{\rm iso}+1)\left( \hat{\rho}_{S,I}(t)\hat{\sigma}_{lc}^+\hat{\sigma}_{lc}^{-} - \hat{\sigma}_{lc}^{-}\hat{\rho}_{S,I}(t)\hat{\sigma}_{lc}^+ \right)\Big] 
\end{equation}
and
\begin{equation}\label{eq:1.40a2}
- \pi J_{l,\rm anis}(\omega_{lc})\Big[\bar{n}_{\rm anis}\left(\hat{\sigma}_{lc}^-\hat{\sigma}_{lc}^+\hat{\rho}_{S,I}(t)-\hat{\sigma}_{lc}^+\hat{\rho}_{S,I}(t)\hat{\sigma}_{lc}^-\right) + (\bar{n}_{\rm anis}+1)\left( \hat{\rho}_{S,I}(t)\hat{\sigma}_{lc}^+\hat{\sigma}_{lc}^{-} - \hat{\sigma}_{lc}^{-}\hat{\rho}_{S,I}(t)\hat{\sigma}_{lc}^+ \right)\Big].
\end{equation}
Instead, the crossing term takes the form
\begin{eqnarray}\label{eq:1.40b}
&& - \pi \sqrt{J_{a,\rm iso}(\omega_{ac})J_{b,\rm iso}(\omega_{ac})}
\Big[\bar{n}_{\rm iso}\left(\hat{\rho}_{S,I}(t)\hat{\sigma}_{bc}^-\hat{\sigma}_{ac}^+ - \hat{\sigma}_{ac}^+\hat{\rho}_{S,I}(t)\hat{\sigma}_{bc}^-\right)\nonumber \\
&& +(\bar{n}_{\rm iso}+1)\left(\hat{\sigma}_{ac}^+\hat{\sigma}_{bc}^-\hat{\rho}_{S,I}(t) - \hat{\sigma}_{bc}^-\hat{\rho}_{S,I}(t)\hat{\sigma}_{ac}^+\right)\Big] 
\end{eqnarray}
and
\begin{eqnarray}\label{eq:1.40c}
&& - \pi \sqrt{J_{a,\rm anis}(\omega_{ac})J_{b,\rm anis}(\omega_{ac})}
\Big[\bar{n}_{\rm anis}\left(\hat{\rho}_{S,I}(t)\hat{\sigma}_{bc}^-\hat{\sigma}_{ac}^+ - \hat{\sigma}_{ac}^+\hat{\rho}_{S,I}(t)\hat{\sigma}_{bc}^-\right)\nonumber \\
&& +(\bar{n}_{\rm anis}+1)\left(\hat{\sigma}_{ac}^+\hat{\sigma}_{bc}^-\hat{\rho}_{S,I}(t) - \hat{\sigma}_{bc}^-\hat{\rho}_{S,I}(t)\hat{\sigma}_{ac}^+\right)\Big].
\end{eqnarray}
In Eqs.~\eqref{eq:1.40a}-\eqref{eq:1.40c}, the labels `${\rm iso}$' and `${\rm anis}$' denote the interaction with the isotropic vacuum reservoir and with a directional excitation reservoir, respectively. The spectral density associated with the isotropic reservoir is
\begin{equation}\label{eq:1.40d}
J_{l,\rm iso}(\omega_{lc}) = \frac{\omega_{lc}^3}{6\hbar\pi^2\varepsilon_{o} c^3}\left|\bm{\mu}_{lc}\right|^{2},
\end{equation}
which we derived in Eq.~\eqref{eq:1.31e}. Instead, the term $\sqrt{J_{a,\rm iso}(\omega_{ac})J_{b,\rm iso}(\omega_{ac})}$ is determined by:
\begin{equation}
\sqrt{J_{a,\rm iso}(\omega_{ac})J_{b,\rm iso}(\omega_{bc})}
= \frac{\omega_{ac}^3}{6\hbar\pi^2\varepsilon_oc^3}\pi\left(\bm{\mu}_{ac}\cdot\bm{\mu}_{bc}\right);
\end{equation}
see also Eq.~\eqref{eq:1.31f}. Moreover, the spectral density of the directional excitation reservoir is
\begin{equation}\label{eq:1.40e}
  J_{l,\rm anis}(\omega_{lc})=\frac{\omega_{lc}^3}{16\hbar\pi^3\varepsilon_0 c^3}
  \sum_{\lambda=1,2} \left|\bm{\mu}_{lc}\cdot\bm{\epsilon}_{\bf{k},\lambda}\right|^{2},
\end{equation}
in agreement with Eq.~\eqref{eq:1.31i}, reflecting the restricted angular support of the driving field's modes.

On the other hand, the average photon occupation numbers $\bar{n}_{\rm iso}$ and $\bar{n}_{\rm anis}$, associated to the two effective reservoirs, are formally given by: 
\begin{equation}
\bar{n}_{\rm iso} = \frac{1}{\exp\left(\frac{\hbar(\omega_{ac}+\omega_{bc})}{2k_B T_{\rm iso}}\right)-1}\
\end{equation}
and
\begin{equation}
\bar{n}_{\rm anis} = \frac{1}{\exp\left(\frac{\hbar(\omega_{ac}+\omega_{bc})}{2k_B T_{\rm anis}}\right)-1}.
\end{equation}
As the isotropic reservoir is modelled as a vacuum reservoir at $T_{\rm iso} = 0$ K, we can set $\bar{n}_{\rm iso} = 0$. This choice stems from the need to model an interaction with a thermal reservoir that leads exclusively to a spontaneous emission process. In contrast, the directional excitation reservoir, leading to absorption and stimulated emission processes, has a finite temperature $T_{\rm anis} > 0$ K, which entails $\bar{n}_{\rm anis}>0$. In the following, we will denote $\bar{n}_{\rm anis} \equiv \bar{n}$, as this is the only term responsible for incoherent pumping.

Although the spontaneous emission into the modes of the directional reservoir at finite temperature is allowed in principle, its associated decay rate is greatly smaller compared to the spontaneous decay rate induced by the vacuum (reservoir at zero temperature). This fact is due to the restricted angular phase space spanned by the modes of the directional reservoir, originated by the interaction with an anisotropic, polarized incoherent field. We will thus neglect such a contribution to spontaneous emission. Moreover, in the limiting case of maximal anisotropy (studied below), where the external radiation has only a single propagation direction and a fixed polarization, the coupling strength of the system with the effective reservoir at finite temperature is substantially weaker than the one with the vacuum; this affects the value of the corresponding spectral density. As we will show in the following sections, having maximal anisotropy is a key ingredient to allow for the emergence of stationary Fano coherences.

It is now necessary to specify the polarization vectors $\bm{\epsilon}_{{\bf k},\lambda}$ of the radiation field. Without loss of generality, we consider a linearly polarized incoherent field with polarization oriented along the $x$ axis. Hence, the corresponding unit polarization vector is
\begin{equation}\label{eq:1.42a}
\bm{\epsilon}_{{\bf k},\lambda} = \bm{\epsilon}_x = \left[1, 0, 0\right].
\end{equation}
Substituting Eq.~\eqref{eq:1.42a} into the expression of the anisotropic spectral density
$J_{l,\mathrm{anis}}$ of Eq.~\eqref{eq:1.40e}, one obtains
$J_{l,\rm anis}(\omega_{lc})=\omega_{lc}^3|\bm{\mu}_{lc}\cdot\bm{\epsilon}_x|^2 / 16\hbar\pi^3\varepsilon_0 c^3$. Inserting the expression of both $J_{l,\rm iso}(\omega_{lc})$ and $J_{l,\rm anis}(\omega_{lc})$ into Eqs.~\eqref{eq:1.40a}-\eqref{eq:1.40c}, the system's master equation is obtained. In particular, one has:
\begin{eqnarray}\label{eq:1.43a}
&& \frac{\omega_{lc}^3|\bm{\mu}_{lc}\cdot\bm{\epsilon}_{x}|^2}{16\hbar\pi^2\varepsilon_0c^3}  \Big[\bar{n}\left(\hat{\sigma}_{ac}^-\hat{\sigma}_{ac}^+\hat{\rho}_S(t)- \hat{\sigma}_{ac}^+\hat{\rho}_S(t)\hat{\sigma}_{ac}^-\right) +\bar{n}\left(\hat{\rho}_S(t)\hat{\sigma}_{ac}^+\hat{\sigma}_{ac}^- -\hat{\sigma}_{ac}^-\hat{\rho}_S(t)\hat{\sigma}_{ac}^+\right)\Big] \nonumber \\
&& + \frac{\omega_{lc}^3|\bm{\mu}_{lc}|^2}{6\hbar\pi\varepsilon_0c^3}\left(\hat{\rho}_S(t)\hat{\sigma}_{ac}^+\hat{\sigma}_{ac}^- -\hat{\sigma}_{ac}^-\hat{\rho}_S(t)\hat{\sigma}_{ac}^+\right),
\end{eqnarray}
which is the contribution ascribable to the diagonal (population) terms of the master equation. The first line of Eq.~\eqref{eq:1.43a} describes absorption and stimulated emission processes induced by the interaction with the anisotropic radiation reservoir, while the second line accounts for the spontaneous emission arising from the coupling to the isotropic vacuum reservoir. Instead, concerning the coherence-generating crossing terms of the system's master equation, we find:
\begin{eqnarray}\label{eq:1.43b}
&& \frac{\omega_{ac}^3\left(\bm{\mu}_{ac}\cdot\bm{\epsilon}_{x}\right)\left(\bm{\mu}_{bc}\cdot\bm{\epsilon}_{x}\right)}{16\hbar\pi^2\varepsilon_0c^3} 
\Big[ \bar{n}\left(\hat{\rho}_S(t)\hat{\sigma}_{bc}^-\hat{\sigma}_{ac}^{+} 
- \hat{\sigma}_{ac}^+\hat{\rho}_S(t)\hat{\sigma}_{bc}^-\right) \nonumber \\
&& 
+ \bar{n}\left(\hat{\sigma}_{ac}^+\hat{\sigma}_{bc}^-\hat{\rho}_S(t) - \hat{\sigma}_{bc}^-\hat{\rho}_S(t)\hat{\sigma}_{ac}^+\right) \Big]
+ \frac{\omega_{ac}^3\left(\bm{\mu}_{ac}\cdot\bm{\mu}_{bc}\right)}{6\hbar\pi\varepsilon_0c^3}\left(\hat{\sigma}_{ac}^+\hat{\sigma}_{bc}^-\hat{\rho}_S(t) - \hat{\sigma}_{bc}^{-}\hat{\rho}_S(t)\hat{\sigma}_{ac}^+\right),\nonumber \\
&&
\end{eqnarray}
where, as before, the first term of the equation in the square brackets originates from the anisotropic driving field, while the second term rules the isotropic spontaneous emission.

In conclusion, the master equation for the interaction of the system with an anisotropic and polarized incoherent radiation field, assumes the form:
{\small
\begin{eqnarray}\label{eq:1.44}
&&\frac{d\hat{\rho}_{S,I}(t)}{dt} = \left.\frac{d\hat{\rho}_{S,I}(t)}{dt}\right|_{\text{incoh}} =\nonumber \\ 
&& - \frac{1}{2}\Big[ \gamma_a^{\rm pol}\bar{n}\big(\hat{\sigma}_{ac}^-\hat{\sigma}_{ac}^+\hat{\rho}_{S,I}(t)-\hat{\sigma}_{ac}^+\hat{\rho}_{S,I}(t)\hat{\sigma}_{ac}^-\big)+(\gamma_a^{\rm pol}\bar{n}+\gamma_a^{\rm iso})\big(\hat{\rho}_{S,I}(t)\hat{\sigma}_{ac}^+\hat{\sigma}_{ac}^- + 
 - \hat{\sigma}_{ac}^-\hat{\rho}_{S,I}(t)\hat{\sigma}_{ac}^+\big) \Big] \nonumber \\
&& - \frac{1}{2}\Big[ \gamma_b^{\rm pol}\bar{n}\big(\hat{\sigma}_{bc}^-\hat{\sigma}_{bc}^+\hat{\rho}_{S,I}(t)-\hat{\sigma}_{bc}^+\hat{\rho}_{S,I}(t)\hat{\sigma}_{bc}^-\big)+(\gamma_b^{\rm pol}\bar{n}+\gamma_b^{\rm iso})\big(\hat{\rho}_{S,I}(t)\hat{\sigma}_{bc}^+\hat{\sigma}_{bc}^- - \hat{\sigma}_{bc}^-\hat{\rho}_{S,I}(t)\hat{\sigma}_{bc}^+\big) \Big] \nonumber \\ 
&& - \frac{1}{2} \sqrt{\gamma_a^{\rm pol}\gamma_b^{\rm pol}}\bar{n}\big(\hat{\rho}_{S,I}(t)\hat{\sigma}_{bc}^-\hat{\sigma}_{ac}^+ - \hat{\sigma}_{ac}^+\hat{\rho}_{S,I}(t)\hat{\sigma}_{bc}^- \big) \nonumber \\
&& -\frac{1}{2}
\left(\sqrt{\gamma_a^{\rm pol}\gamma_b^{\rm pol}}\bar{n}+p\sqrt{\gamma_a^{\rm iso}\gamma_b^{\rm iso}}\right)\big(\hat{\sigma}_{ac}^+\hat{\sigma}_{bc}^-\hat{\rho}_{S,I}(t) - \hat{\sigma}_{bc}^-\hat{\rho}_{S,I}(t)\hat{\sigma}_{ac}^+\big) \nonumber \\
&& - \frac{1}{2} \sqrt{\gamma_a^{\rm pol}\gamma_b^{\rm pol}}\bar{n}\big(\hat{\rho}_{S,I}(t)\hat{\sigma}_{ac}^-\hat{\sigma}_{bc}^+ - \hat{\sigma}_{bc}^+\hat{\rho}_{S,I}(t)\hat{\sigma}_{ac}^- \big) \nonumber \\
&& -\frac{1}{2}
\left(\sqrt{\gamma_a^{\rm pol}\gamma_b^{\rm pol}}\bar{n}+p\sqrt{\gamma_a^{\rm iso}\gamma_b^{\rm iso}}\right)\big(\hat{\sigma}_{bc}^+\hat{\sigma}_{ac}^-\hat{\rho}_{S,I}(t) - \hat{\sigma}_{ac}^-\hat{\rho}_{S,I}(t)\hat{\sigma}_{bc}^+\big)+\rm{h.c.}.
\end{eqnarray}
}
where the following definitions have been used: 
\begin{eqnarray}
\gamma_l^{\rm iso} &\equiv& \frac{\omega_{lc}^3\left|\bm \mu_{lc}\right|^2}{3\hbar\pi \varepsilon_0 c^3} \label{eq:1.45a}\\ 
p &\equiv& \frac{\bm{\mu}_{ac}\cdot\bm{\mu}_{bc}}{\left|\bm{\mu}_{ac}\right|\left|\bm{\mu}_{bc}\right|}=\cos(\Theta) \label{eq:1.45b}\\
\gamma_l^{\rm pol} &\equiv&\frac{\omega_{lc}^{3} |\bm{\mu}_{lc}\cdot\bm{\epsilon}_{x}|^2}{8\hbar\pi^2 \varepsilon_0 c^3}.\label{eq:1.45c}
\end{eqnarray}
The quantity $\gamma_l^{\mathrm{iso}}$ denotes the \emph{spontaneous decay rate} from the excited state $|l\rangle$ $(l=a,b)$ to the common ground state $|c\rangle$. The parameter $p$ is the \emph{alignment parameter} between the dipole moment vectors $\bm \mu_{ac}$ and $\bm \mu_{bc}$ of the energy transitions, and is univocally defined through the angle $\Theta$ between them. The range of values for $p$ is from $-1$ to $+1$, where $p = +1$ indicates that the dipoles $\bm{\mu}_{ac}$ and $\bm{\mu}_{bc}$ are parallel, $p = -1$ that they are anti-parallel, and $p = 0$ that they are mutually orthogonal. When $p=0$, the interference terms proportional to $p\sqrt{\gamma_a^{\rm iso}\gamma_b^{\rm iso}}$ vanish. Finally, the quantities $r_l^{\rm pol} \equiv \bar{n}\gamma_l^{\rm pol}$ describe the incoherent pumping rates associated with \emph{absorption} and \emph{stimulated emission} driven by the polarized input field.

Finally, we derive the equation of motion for $\hat{\rho}_S(t)$ in the Schr\"odinger picture by adding the Hamiltonian $\hat{H}_S$ of the three-level system in the coherent part of the differential equation of $\hat{\rho}_S(t)$. It entails to solve the differential equation
\begin{eqnarray}\label{eq:1.46}
\frac{d\hat{\rho}_S(t)}{dt} &=& -\frac{i}{\hbar}{\rm Tr}_R\left[ \hat{H}_I(t) + \hat{H}_S \otimes \hat{I}_R, \hat{\rho}_S(0) \otimes \hat{\rho}_R(0) \right]\nonumber \\
&-& \frac{1}{\hbar^2}\int_0^t {\rm Tr}_R\left[ \hat{H}_I(t),\left[\hat{H}_I(t'),\hat{\rho}_S(t')\otimes\hat{\rho}_R(0)\right]\right]dt',
\end{eqnarray}
where $\hat{I}_R$ denotes the identity operator defined in the reservoir's Hilbert space. In this way, decomposing $\hat{\rho}_S(t)$ in its elements $\langle l|\hat{\rho}_S(t)|j\rangle \equiv \rho_{lj}(t)$ with $l,j=a,b,c$, the set of differential equations for each $\rho_{lj}(t)$ is obtained:
{\small
\begin{equation}\label{eq:1.47a}
\begin{cases}
\displaystyle{ \frac{ d\rho_{aa}(t) }{dt} = - \left(\gamma_a^{\rm pol}\bar{n} + \gamma_a^{\rm iso}\right)\rho_{aa}(t)  + \gamma_a^{\rm pol} \bar{n} \rho_{cc}(t) -\left(\sqrt{\gamma_a^{\rm pol} \gamma_b^{\rm pol}}\bar{n} + p\sqrt{\gamma_a^{\rm iso} \gamma_b^{\rm iso}} \right)\rm{Re}[\rho_{ab}(t)] }\\
\displaystyle{ \frac{ d\rho_{bb}(t) }{dt} = - \left(\gamma_b^{\rm pol}\bar{n} + \gamma_b^{\rm iso}\right)\rho_{bb}(t)  + \gamma_b^{\rm pol} \bar{n} \rho_{cc}(t) -\left(\sqrt{\gamma_a^{\rm pol} \gamma_b^{\rm pol}}\bar{n} + p\sqrt{\gamma_a^{\rm iso} \gamma_b^{\rm iso}} \right)\rm{Re}[\rho_{ab}(t)]}\\
\displaystyle{ 
\begin{split}\frac{ d\rho_{cc}(t) }{dt} =& -\left(\gamma_a^{\rm pol}+\gamma_b^{\rm pol}\right)\bar{n}\rho_{cc}(t) + \bar{n} \left( \gamma_a^{\rm pol}\rho_{aa}(t) + \gamma_b^{\rm pol} \rho_{bb}(t) \right) + \left( \gamma_a^{\rm iso}\rho_{aa}(t) + \gamma_b^{\rm iso}\rho_{bb}(t) \right) \\
&+ 2\left(\sqrt{\gamma_a^{\rm pol} \gamma_b^{\rm pol}}\bar{n} + p\sqrt{\gamma_a^{\rm iso} \gamma_b^{\rm iso}} \right)\rm{Re}[\rho_{ab}(t)]
\end{split}}\\
\displaystyle{
\begin{split} \frac{ d\rho_{ab}(t) }{dt} =& -\frac{1}{2}\left(\sqrt{\gamma_a^{\rm pol} \gamma_b^{\rm pol}}\bar{n} + p\sqrt{\gamma_a^{\rm iso} \gamma_b^{\rm iso}}\right)\left(\rho_{aa}(t)+\rho_{bb}(t)\right) + \sqrt{\gamma_a^{\rm pol} \gamma_b^{\rm pol} }\bar{n}\rho_{cc}(t)\\
& - \left[\frac{\gamma_a^{\rm pol}+\gamma_b^{\rm pol}}{2}\bar{n} +\frac{\gamma_a^{\rm iso}+\gamma_b^{\rm iso}}{2}+i\rm{\Delta}\right]\rho_{ab}(t),
\end{split} }
\end{cases}
\end{equation}
}
together with
{\small
\begin{equation}\label{eq:1.47b}
\begin{cases}
\displaystyle{
\begin{split} \frac{ d\rho_{ac}(t) }{dt} = & -\frac{1}{2}\left(\sqrt{\gamma_a^{\rm pol} \gamma_b^{\rm pol} }\bar{n} + p\sqrt{\gamma_a^{\rm iso} \gamma_b^{\rm iso} }\right)\rho_{bc}(t) -\left[\frac{\gamma_b^{\rm pol}}{2}\bar{n}+\gamma_a^{\rm pol}\bar{n}+\frac{\gamma_a^{\rm iso}}{2} + i\frac{\omega_{ac}}{2}\right]\rho_{ac}(t) 
\end{split}} \\
\displaystyle{
\begin{split} \frac{ d\rho_{bc}(t) }{dt} = & -\frac{1}{2}\left(\sqrt{\gamma_a^{\rm pol} \gamma_b^{\rm pol} }\bar{n} + p\sqrt{\gamma_a^{\rm iso} \gamma_b^{\rm iso} }\right)\rho_{ac}(t) -\left[\frac{\gamma_a^{\rm pol}}{2}\bar{n}+\gamma_b^{\rm pol}\bar{n}+\frac{\gamma_b^{\rm iso}}{2} + i\frac{\omega_{bc}}{2}\right]\rho_{bc}(t) .
\end{split}}
\end{cases}
\end{equation}
}

Eqs.~(\ref{eq:1.47a})-(\ref{eq:1.47b}) correspond to two independent dynamical sub-processes of the quantum system~\cite{Agarwal}. Eq.~(\ref{eq:1.47a}) describes the time evolution of the populations and coherence between the two nearly degenerate excited states $|a\rangle$ and $|b\rangle$ of the V-type system. This coherence arises from the interference between the radiative pathways connecting each excited level to the common ground state $|c\rangle$, occurring both in absorption and in spontaneous or stimulated emission (see Fig.~\ref{fig:1})
Even though the direct optical transition 
$|a\rangle\leftrightarrow|b\rangle$ is dipole forbidden, coherence can nevertheless be generated indirectly due to the shared coupling to  $|c\rangle$. When the radiation field is sufficiently broadband, such that its spectral width exceeds the excited state splitting, the transitions $|a\rangle\leftrightarrow|c\rangle$ and $|b\rangle\leftrightarrow|c\rangle$   
become indistinguishable. As a consequence, the environment can not determine which excitation or decay channel has been involved, giving rise to interference between the two optical pathways and allowing for the generation of Fano coherence.
In contrast, the sub-process (\ref{eq:1.47b}) returns the time evolution of the one-photon coherences $\rho_{ac}$ and $\rho_{bc}$, which decouple from the populations dynamics. This decoupling follows from the partial secular approximation, where fast oscillations at the optical transition frequencies are averaged out while retaining terms oscillating at the excited state splitting $\Delta$~\cite{Koyu-Dodin,Donati}.

Albeit the structure of Eqs.~\eqref{eq:1.47a}-\eqref{eq:1.47b} is formally analogous to that obtained for isotropic radiation (see Refs.~\cite{Dodin-Tscherbul,Kozlov,Tscherbul-Brumer,Donati}), the present configuration involves two distinct reservoirs, characterised by the rates $\gamma_l^{\rm pol}$ and $\gamma_l^{\rm iso}$. The subset of field modes with wavevector ${\bf k}$ and polarization $\bm{\epsilon}_x$ forms an effective hot reservoir ($T_{\rm anis}>0K$) that drives absorption and stimulated emission, whereas the remaining modes constitute a cold reservoir ($T_{\rm iso}=0K$) responsible for spontaneous decay, as previously discussed~\cite{Dodin-Brumer}.  
When $\gamma_l^{\mathrm{iso}}\neq 0$ the V-system is effectively \emph{open}, since spontaneous emission may populate modes or levels lying outside the selected three-level system. Physically, the polarized excitation field selects only a restricted set of radiation modes (and therefore a subset of allowed transitions), while the residual spontaneous emission channels, allowed by selection rules but not addressed by the driving configuration, are described by $\gamma_l^{\mathrm{iso}}$. In experimental realizations with atoms (refer also to Subsec.~\ref{subsec:3.1}), population loss from the targeted V-manifold can be mitigated using mechanisms that re-distribute population among Zeeman sublevels. 
For instance, in the presence of a transverse magnetic field, Zeeman precession within the $F=1$ manifold induces population mixing among the magnetic sublevels, thereby repopulating the subspace spanned by the V-type system and effectively restoring closed-system behaviour on the timescale of interest.

To illustrate these concepts, we now consider a physically relevant example, where we select two specific dipole moment vectors $\bm{\mu}_{ac}$ and $\bm{\mu}_{bc}$ associated with the energy transitions $|a\rangle \leftrightarrow |c\rangle$ and $|b\rangle \leftrightarrow |c\rangle$. We assume that the transitions couple to circularly polarized radiation in the $x$-$y$ plane travelling along the $z$ axis, a common configuration in atomic systems. Left- and right-handed circular polarization transitions correspond to the unit dipole vectors: 
\begin{equation}\label{eq:1.41a}
\bm{\mu} = \bm{\mu}_+ = \left[-\frac{1}{\sqrt{2}}, -i\frac{1}{\sqrt{2}}, 0\right]
\end{equation}
and
\begin{equation}\label{eq:1.41b}
\bm{\mu} = \bm{\mu}_- = \left[\frac{1}{\sqrt{2}}, -i\frac{1}{\sqrt{2}}, 0\right],
\end{equation}
respectively.
We assume that the dipole moment vector $\bm{\mu}_{ac}$ for the transition $|a\rangle\leftrightarrow|c\rangle$ is $\bm{\mu}_{ac}=|\bm{\mu}_{ac}|{\bm\mu}_-$, and the dipole moment vector $\bm{\mu}_{bc}$ for the transition $|b\rangle\leftrightarrow|c\rangle$ is $\bm{\mu}_{bc}=|\bm{\mu}_{bc}|{\bm\mu}_+$. It is worth noting that $\bm{\mu}_{ac}\,\bot\,\bm{\mu}_{bc}$. A linearly polarized radiation along the $x$-axis ($\bm{\epsilon}_{\lambda} = \bm{\epsilon}_x$), as given by Eq.~\eqref{eq:1.42a}, can be rewritten as
\begin{equation}\label{eq:1.42b}
\bm{\epsilon}_x = \sqrt{2}\left(\bm{\mu}_{-} - \bm{\mu}_+\right).
\end{equation}
This means that a radiation with polarization $\bm{\epsilon}_x $ can drive both transitions $|a\rangle \leftrightarrow |c\rangle$ and $|b\rangle \leftrightarrow |c\rangle$.\\
The two selected transitions $|a\rangle\leftrightarrow|c\rangle$ and $|b\rangle\leftrightarrow|c\rangle$ have orthogonal dipole moments, thus one has $p=0$ for the isotropic spontaneous emission channel. Consequently, in the case of isotropic and unpolarized radiation, the interference terms vanish and the dynamical equations reduce to the standard Pauli rate equations~\cite{Dodin-Brumer,Koyu-Tscherbul}.
In contrast, polarized excitation preserves interference in the absorption and stimulated emission processes, allowing the coherence $\rho_{ab}$ to persist even for orthogonal dipole moments, due to the anisotropic pumping.\\
Finally, we show that the decay rate $\gamma_l^{\rm pol}$ induced by the polarized incoherent field is significantly smaller than the spontaneous decay rate  $\gamma_{l}^{\rm iso}$ associated with the isotropic vacuum. For the configuration considered here, $\gamma_{l}^{\rm pol}$ reads: 
\begin{equation}\label{eq:1.48}
  \gamma_l^{\rm pol}  = \frac{\omega_{lc}^3\left|\mu_{lc}\right|^2}{16\hbar\pi^2 \varepsilon_0 c^3} = \gamma_l^{\rm iso} \frac{3}{16 \pi}\quad \Longrightarrow \quad \gamma_l^{\rm pol} = \frac{3}{16 \pi}\gamma_l^{\rm iso} \nonumber.
\end{equation}
This result shows that the decay rate induced by polarized radiation is reduced by a factor  $16\pi/3$ with respect to the isotropic spontaneous emission rate~\cite{Koyu-Dodin}. These expressions apply to a closed V-type system, where all spontaneous decay channels repopulate the ground state $|c\rangle$.

%%%%%%%%%%%%%%%%%%%%%%%%%%%%%%%%
\subsubsection{Open system dynamics in linear form}

Eqs.~(\ref{eq:1.47a})-(\ref{eq:1.47b}) model the dynamics of a V-type three level system interacting with a continuum of radiation modes, with wavevector $\bf k$ and polarization $\bm{\epsilon}_x$. As previously discussed, these equations refer to two independent sub-processes. Therefore, the solution can be achieved by solving two distinct systems of {\it linear} equations, i.e., 
\begin{equation}\label{eq:1.49a}
\frac{ d{\bf{x}}(t) }{dt} = A{\bf x}(t) \quad \text{and} \quad
\frac{ d{\bf{z}}(t) }{dt} = C{\bf z}(t)
\end{equation}
with state vectors
\begin{eqnarray}\label{eq:1.49b}
{\bf x}(t) &\equiv& \Big( \rho_{aa}(t), \rho_{bb}(t), \rho_{cc}(t), {\rm Re}[\rho_{ab}(t)], {\rm Im}[\rho_{ab}(t)]\Big)^T \\
{\bf z}(t) &\equiv& \Big( {\rm Re}[\rho_{ac}(t)], {\rm Im}[\rho_{ac}(t)], {\rm Re}[\rho_{bc}(t)], {\rm Im}[\rho_{bc}(t)]\Big)^{T}.
\end{eqnarray}
The vector ${\bf x}$ includes the population of the ground level $\rho_{cc}(t)$ rather than imposing the constraint $\rho_{cc}(t) = 1 - \rho_{aa}(t) - \rho_{bb}(t)$~\cite{Donati}. This choice is needed to get at any time $t$ the correct density operator $\hat{\rho}_S(t)$, solution of Eqs.~\eqref{eq:1.47a}-\eqref{eq:1.47b} altogether, from the direct exponentiation of the two differential equations in \eqref{eq:1.49a}. In Eq.~\eqref{eq:1.49a}, the $ij$-th (row $i$ and column $j$) elements of the matrices $A,C$ are respectively:
\begin{eqnarray}
    && A_{11} = -A_{31} = -\left(\gamma_{a}^{\rm pol}\bar{n}+\gamma_{a}^{\rm iso}\right)\nonumber\\ 
    && A_{12} = A_{51} = A_{21} = A_{52} = A_{53} = A_{15} = A_{25} = A_{35} = 0\nonumber\\ 
    && A_{41} = A_{42} = \frac{ A_{14} }{2} = \frac{ A_{24} }{2} = - \frac{ A_{34} }{4} = -\frac{1}{2}\left(\sqrt{\gamma_a^{\rm pol}\gamma_b^{\rm pol}}\bar{n}+p\sqrt{\gamma_a^{\rm iso}\gamma_b^{\rm iso}}\right)\nonumber\\ 
    && A_{22} = -A_{32} = -\left(\gamma_{b}^{\rm pol}\bar{n}+\gamma_{b}^{\rm iso}\right)\nonumber\\
    && A_{13} = \gamma_{a}^{\rm pol}\bar{n}\nonumber\\
    && A_{23} = \gamma_{b}^{\rm pol}\bar{n}\nonumber\\
    && A_{33} = \left(\gamma_{a}^{\rm pol}+\gamma_{b}^{\rm pol}\right)\bar{n}\nonumber\\
    && A_{43} = \sqrt{\gamma_a^{\rm pol}\gamma_b^{\rm pol}}\bar{n}\nonumber\\
    && A_{44} = A_{55} = \left(\frac{\gamma_a^{\rm pol}+\gamma_b^{\rm pol}}{2}\bar{n}+\frac{\gamma_a^{\rm iso}+\gamma_b^{\rm iso}}{2}\right)\nonumber\\
    && A_{54} = -A_{45} = -\Delta,
\end{eqnarray}
and
\begin{eqnarray}
    && C_{11} = C_{22} = - \left[\bar{n}\left(\gamma_a^{\rm pol}+\frac{\gamma_b^{\rm pol}}{2}\right)+\frac{\gamma_a^{\rm iso}}{2}\right]\nonumber\\
    && C_{21} = -C_{12} = -\omega_{ac};\nonumber\\
    && C_{31} = C_{42} = C_{13} = C_{24} = -\frac{1}{2}\left(\sqrt{\gamma_a^{\rm pol}\gamma_b^{\rm pol}}\bar{n}+p\sqrt{\gamma_a^{\rm iso}\gamma_b^{\rm iso}}\right)\nonumber\\
    && C_{41} = C_{32} = C_{23} = C_{14} = 0;\nonumber\\
    && C_{33} = C_{44} = - \left[\bar{n}\left(\gamma_b^{\rm pol}+\frac{\gamma_a^{\rm pol}}{2}\right)+\frac{\gamma_b^{\rm iso}}{2}\right]\nonumber\\
    && C_{43} = -C_{34} = \omega_{bc}. 
\end{eqnarray}

The homogeneous differential equations (\ref{eq:1.49a}) are solved numerically via exponentiation, namely
\begin{eqnarray}
    {\bf x}(t) &=& e^{At}{\bf x}(0)\label{eq:sol_x_t} \\
    {\bf z}(t) &=& e^{Ct}{\bf z}(0)\label{eq:sol_z_t}
\end{eqnarray}
with ${\bf x}(0),\,{\bf z}(0)$ denoting the initial states in this representation. The exponential of the matrices $A,C$ is computed using the \textsc{Matlab} function \texttt{expm}, which exploits the scaling and squaring algorithm of Higham~\cite{Higham}. 

\subsection{Dynamical regimes and Fano coherence lifetime}\label{subsec:2.3}

In order to analyse the generation of Fano coherence under incoherent driving, we start from the condition where the system is initialized in a coherence-free state. Following Refs.~\cite{Koyu-Tscherbul,Koyu-Dodin,Dodin-Tscherbul,Tscherbul-Brumer2}, we choose the ground state as the initial condition, so that the state vector ${\bf x}(0)$ contains only one non-zero entry, $\rho_{cc}(0)=1$. This guarantees that the density operator, solution of the dynamics, remains semi-positive definite at all times, as explicitly shown in Refs.~\cite{Koyu-Tscherbul,Dodin-Tscherbul,Tscherbul-Brumer2}.

Analytical solutions of Eqs.~\eqref{eq:1.47a}-\eqref{eq:1.47b} have been obtained in Refs.~\cite{Dodin-Brumer,Koyu-Dodin}, where the dynamics of Fano coherence $\rho_{ab}(t)$ fall into three distinct physical regimes. These regimes are controlled by the ratio $\Delta/\bar{\gamma}$, where $\Delta$ is the excited state energy splitting and $\bar{\gamma}=(\gamma_{a}^{\mathrm{iso}}+\gamma_{b}^{\mathrm{iso}})/2$ is the average spontaneous emission rate. In particular, $\Delta/\bar{\gamma} \ll 1$ corresponds to {\it overdamping}, while $\Delta/\bar{\gamma}\gg 1$ to {\it underdamping}, where $\Delta/\bar{\gamma}\approx 1$ defines the transition separating the two. The behaviour in each regime also depends on the average photon number $\bar{n}$ of the driving field, which quantifies the radiation intensity. This allows one to distinguish different driving conditions, commonly referred to as the weak- and strong-pumping regimes. Without loss of generality, we assume $\gamma_a^{\mathrm{iso}} \geq \gamma_b^{\mathrm{iso}}$, although all results remain valid for the opposite condition.

The regime structure and numerical behaviour discussed below follow the results of Dodin \emph{et al.}~\cite{Dodin-Brumer} and Koyu \emph{et al.}~\cite{Koyu-Dodin}.

\subsubsection{Weak-pumping regime}

Let us first consider the \emph{weak-pumping regime}, defined by $\bar{n}\ll 1$, which corresponds to low radiation power that is representative, for example, of non-concentrated solar illumination. In this regime, we examine both symmetric ($\gamma_a^{\mathrm{iso}}=\gamma_b^{\mathrm{iso}}$) and asymmetric ($\gamma_a^{\mathrm{iso}} > \gamma_b^{\mathrm{iso}}$; we will consider $\gamma_a^{\mathrm{iso}} = 10 \gamma_b^{\mathrm{iso}}$) spontaneous decay configurations. 

\begin{figure}[t]
\centering
\includegraphics[width=0.75\textwidth]{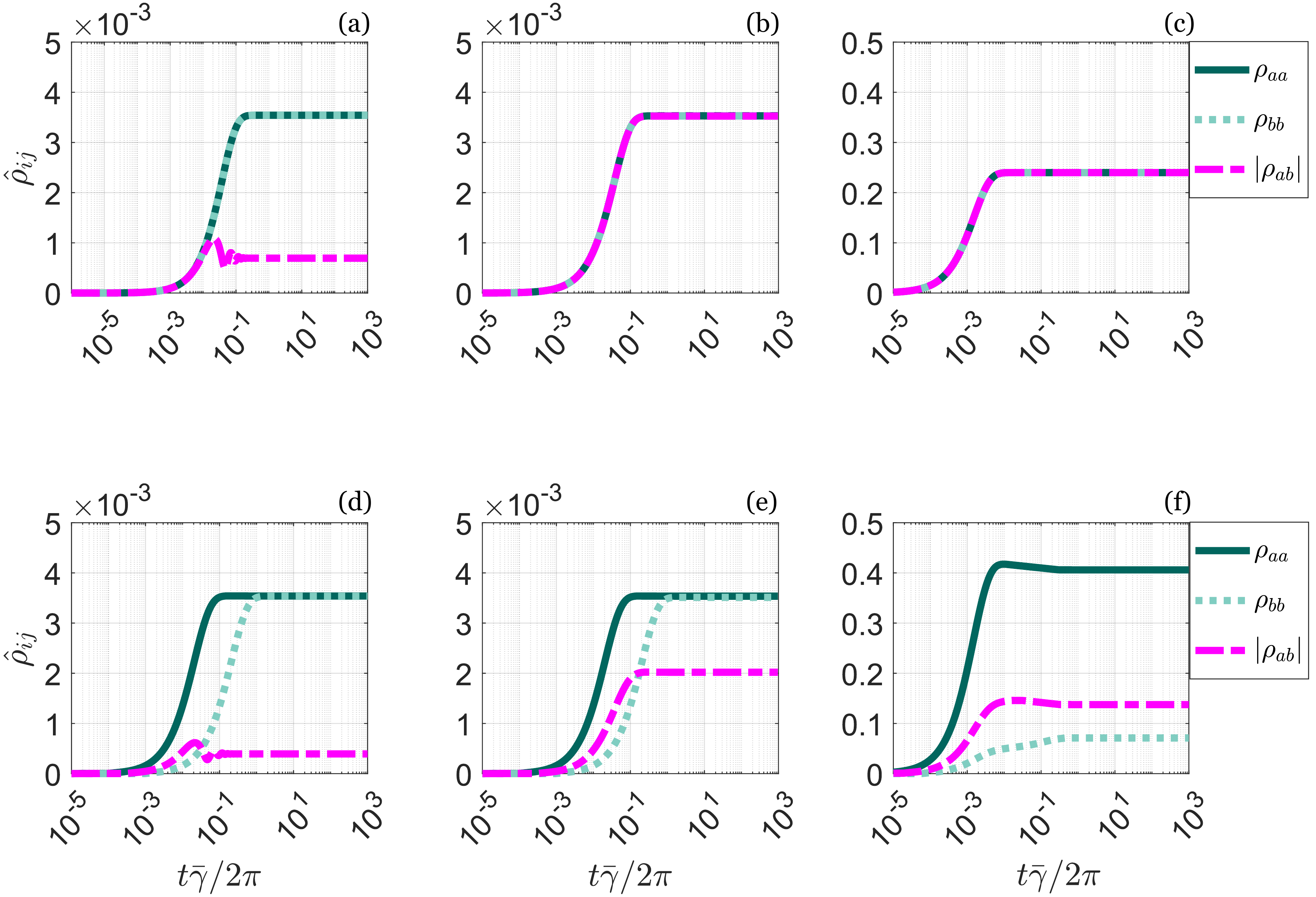}
\caption{Dimensionless time evolution of both the excited levels populations and module of quantum coherence $|\rho_{ab}|$. (a):  $\gamma_{a}^{\rm iso}/\gamma_{b}^{\rm iso}=1$, $\Delta/\bar{\gamma}=10$, $\bar{n}=0.06$; (b): $\gamma_{a}^{\rm iso}/\gamma_{b}^{\rm iso}=1$, $\Delta/\bar{\gamma}=0.1$, $\bar{n}=0.06$; (c): $\gamma_{a}^{\rm iso}/\gamma_{b}^{\rm iso}=1$, $\Delta/\bar{\gamma}=0.1$, $\bar{n}=100$; (d): $\gamma_{a}^{\rm iso}/\gamma_{b}^{\rm iso}=10$, $\Delta/\bar{\gamma}=10$, $\bar{n}=0.06$; (e): $\gamma_{a}^{\rm iso}/\gamma_{b}^{\rm iso}=10$, $\Delta/\bar{\gamma}=0.1$, $\bar{n}=0.06$; (f): $\gamma_{a}^{\rm iso}/\gamma_{b}^{\rm iso}=10$, $\Delta/\bar{\gamma}=0.1$, $\bar{n}=100$. The panels in the first row refer to symmetric systems, while the panels in the second row are related to asymmetric ones. 
}
\label{fig:2}
\end{figure}

\begin{itemize}
\item [(i)] {\bf Underdamping, $\Delta/\bar{\gamma}\gg 1$}:\\
The real and imaginary parts of the coherence $\rho_{ab}$ oscillate over time at frequency $\Delta$, approaching a stationary state on the timescale $\tau_{\rm coh}\approx 1/\bar{\gamma}$. Remarkably, the steady-state magnitude $|\rho_{ab}|_{\rm steady}$ is non-zero, indicating that Fano coherences do not decay under polarized incoherent driving. This behaviour occurs in both symmetric and asymmetric systems, as shown in Fig.~\ref{fig:2}(a) and \ref{fig:2}(d). Although counter-intuitive, the persistence of coherence does not violate thermodynamic principles: as argued in Ref.~\cite{Dodin-Brumer}, polarized radiation effectively acts as a high-temperature reservoir coupled to both transitions $|a\rangle \leftrightarrow |c\rangle$ and $|b\rangle \leftrightarrow |c\rangle$, while spontaneous emission is isotropic and can be modelled by the effective interaction of the system with also a cold reservoir. The simultaneous interaction with these two reservoirs keeps the system out-of-equilibrium, allowing stationary coherence to survive.

In the weak-pumping, underdamped regime, the population dynamics are well captured by the secular approximation. This is primarily due to the large excited-state splitting $\Delta\gg\bar{\gamma}$, which causes rapid coherence oscillations that average out non-secular terms, while weak-pumping further limits the magnitude of coherence generation. 
\end{itemize}

\begin{itemize}
\item [(ii)] {\bf Overdamping, $\Delta/\bar{\gamma}\ll 1$}:

The coherence dynamics differ qualitatively from those of the underdamping case. As illustrated in Fig.~\ref{fig:2}(b) and Fig.~\ref{fig:2}(e) for symmetric and asymmetric systems, respectively, the magnitude of $\rho_{ab}$ evolves toward a non-zero stationary value, with no oscillatory behaviour. Such trend marks a clear departure also from the isotropic and unpolarized scenario, where stationary coherence can persist only in the strictly degenerate limit $\Delta = 0$.

Another notable feature of having input polarized radiation is that, for a symmetric V-system, the coherence magnitude $|\rho_{ab}|$ becomes practically equal to the excited state populations, as derived in Ref.~\cite{Dodin-Brumer}. In asymmetric configurations, the imbalance between decay rates leads to unequal relaxation of the excited states: as illustrated in Fig.~\ref{fig:2}(e), the population $\rho_{bb}(t)$ approaches its steady value more slowly than $\rho_{aa}(t)$. This imbalance reduces the interference responsible for coherence generation, resulting in a smaller stationary value of $|\rho_{ab}|$ compared to the symmetric case.
\end{itemize} 

\subsubsection{Strong-pumping regime}\label{subsec:strong-pumping regime}

In the \emph{strong-pumping} regime, defined by $\bar{n}\gg 1$, overdamping can arise even for significantly larger values of $\Delta/\bar{\gamma}$ than in the weak-pumping regime~\cite{Koyu-Tscherbul}. This broadening of the overdamping domain reflects the relevant role played by stimulated processes at high photon occupation numbers.
In this regime, we set $\bar{n}=100$ and $\Delta/\bar{\gamma}=0.1$. As shown in Fig.~\ref{fig:2}(c) and Fig.~\ref{fig:2}(f) for the symmetric and asymmetric systems respectively, the stationary coherence $|\rho_{ab}|$ reaches much higher values than the ones obtained under weak pumping. 
As clarified by Koyu \emph{et al.} in Ref.~\cite{Koyu-Tscherbul}, the condition 
$\bar{n}\gg 1 $ implies 
$r_l^{\rm pol}/\gamma_l^{\rm pol}\gg 1$ ($l=a,b$) and does not violate the weak-coupling assumption, provided that the incoherent pumping rates satisfy $r_l^{\rm pol}\ll \omega_{lc}$. In other words, the pumping rates must remain much smaller than the corresponding optical transition frequencies. For optical transitions with  $\omega_{lc}\approx 100$ THz
and spontaneous decay rates 
$\gamma_l^{\rm iso}\approx 10$ MHz, the associated polarized decay rates $\gamma_l^{\rm pol} = \frac{3}{16 \pi}\gamma_l^{\rm iso}$
allow average photon numbers $\bar{n} \ll 10^8$ while remaining well within the weak-coupling regime.
Regarding the Markovianity of the system's dynamics, it is maintained if the pumping rate $r_l^{\rm pol}$ is smaller than the inverse of the reservoir correlation time $\tau_R$. Typically, in the optical regime $\tau_R\approx 10$ fs; thus, $\bar{n} \ll 10^8$ is acceptable. 
In the following, however, we restrict our analysis to values up to $\bar{n} \approx 10^4$, which lie well within this validity range and provide a conservative regime where all approximations employed in the derivation of the master equation are safely satisfied.

\begin{figure}[t]
\centering
\includegraphics[width=0.75\textwidth]{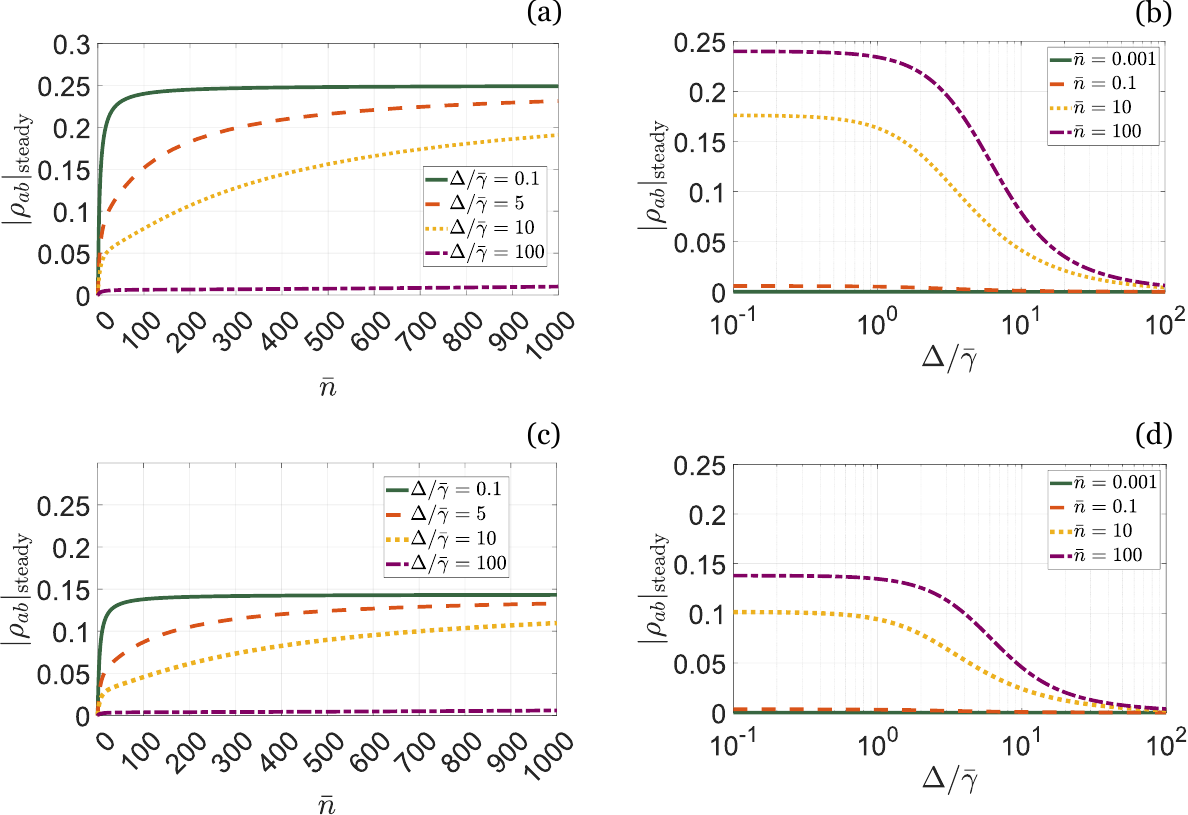}
\caption{
Excited state coherence $\rho_{ab}$ in the long-time limit $t\rightarrow\infty$ as a function of $\bar{n}$ and $\Delta/\bar{\gamma}$. (a) Plot of $|\rho_{ab}|_{\rm steady}$ varying $\bar{n}$, for different values of $\Delta/\bar{\gamma}$ in a symmetric system. (b) Plot of $|\rho_{ab}|_{\rm steady}$ varying $\Delta/\bar{\gamma}$, for different values of $\bar{n}$ in a symmetric system. (c) Plot of $|\rho_{ab}|_{\rm steady}$ varying $\bar{n}$, for different values of $\Delta/\bar{\gamma}$ in an asymmetric system ($\gamma_a^{\rm iso}/\gamma_b^{\rm iso}=10$). (d) Plot of $|\rho_{ab}|_{\rm steady}$ varying $\Delta/\bar{\gamma}$, for different values of $\bar{n}$ in an asymmetric system ($\gamma_a^{\rm iso}/\gamma_b^{\rm iso}=10$).
}
\label{fig:3}
\end{figure}  

A large mean photon number $\bar{n}$ combined with a small excited state energy splitting $\Delta$ identifies the region where the stationary coherence $|\rho_{ab}|_{\rm steady}$ reaches its highest values. The dependence of $|\rho_{ab}|_{\rm steady}$ on $\bar{n}$ and $\Delta/\bar{\gamma}$ is illustrated in Fig.~\ref{fig:3}, either for a symmetric V-type system 
($\gamma_a/\gamma_b=1$) and for a strongly asymmetric one ($\gamma_a/\gamma_b=10$).  
In panels (a) and (c), we plot the stationary coherence as a function of $\bar{n}$ for several fixed values of $\Delta/\bar{\gamma}$, whereas in panels (b) and (d) we report $|\rho_{ab}|_{\rm steady}$ versus $\Delta/\bar{\gamma}$ for various fixed $\bar{n}$. Panels (a)-(b) refer to the symmetric systems, and panels (c)-(d) to the asymmetric ones. These complementary plots highlight that decreasing $\Delta$ or increasing $\bar{n}$ enhances the attainable steady-state coherence in both symmetric and asymmetric V-type systems. However, the maximum coherence achievable differs between the symmetric and asymmetric configurations. This difference originates from the positivity constraint of the system's density operator, which imposes $|\rho_{ab}| \leq \sqrt{\rho_{aa}\rho_{bb}}$.
In symmetric systems, the two excited levels acquire comparable stationary populations, allowing $|\rho_{ab}|_{\rm steady}$ to approach its maximal upper bound. In contrast, when the decay rates are strongly unbalanced, the stationary populations become unequal ($\rho_{aa}\neq\rho_{bb}$), and the value $\sqrt{\rho_{aa}\rho_{bb}}$ of the upper bound is reduced. As a result, the maximum coherence attainable in the asymmetric case is necessarily smaller.

Furthermore, if the system starts in the ground state, incoherent pumping can transfer at most one quarter of the total population to each excited level at the steady state, leading to the bound $|\rho_{ab}|_{\rm stat}\leq 0.25$. This fact is in agreement with the behaviours observed in Fig.~\ref{fig:3}(a) and with the analysis in Ref.~\cite{Kozlov}.

It is important to clarify that the solutions presented in this paper are based on the assumption of an instantaneous activation of the interaction between the system and an incoherent radiation field. Dodin \emph{et al} in \cite{Dodin-Tscherbul2} discuss time-dependent incoherent radiation, showing that if the turn-on time exceeds the system's fastest characteristic time scale $\tau_S$, the magnitude of induced Fano coherence can significantly decrease. However, if the incoherent field is realized by a broadband laser, sufficiently rapid activation can be achieved using acousto-optic modulators. 

%%%%%%%%%%%%%%%%%%%%%%%%%%%%%%%%
\section{Steady-state Fano coherence in relevant experimental setups}\label{sec3}

In this section, we aim to report the calculation of Fano coherence at the steady-state for a relevant experimental setup, which could serve as a suitable testing ground for initial verification and subsequent exploitation. In addition, we also discuss the experimental challenges that this setting could present, and possible solutions to overcome or at least mitigate them.

The stationary of Fano coherence under ``noisy'' conditions, necessary for its generation, can be achieved in systems in contact with thermal reservoirs, such as quantum heat engines~\cite{Scully-Chapin}, or systems exposed to thermal radiation as in our case-study and in photo-conversion devices~\cite{Svidzinsky,Scully}. In particular, in the latter case, Svidzinsky \emph{et al.}~\cite{Svidzinsky} theoretically demonstrate that Fano interferences might enable the mitigation of spontaneous emission, thereby reducing radiative recombination phenomena. The photo-conversion device (a photocell) is modelled with a V-type three-level system driven by incoherent light source, wherein the excited states represent conduction band states decaying into a common valence band state. Quantum coherence between the excited states of the system could theoretically lead to an increase in the extractable current from the device. Such an enhancement boosts the output power and conversion efficiency. On the other hand, implementing a ${\rm \Lambda}$-type system can enhance photon absorption, as demonstrated by Scully \emph{et al.} in Ref.~\cite{Scully-Chapin} using a quantum heat engine model. A better photon absorption is allowed by a rapid transfer of atoms from the ground levels to the excited level that enables an increase in the extractable work. The boost in performance for both applications, whether in photocells or quantum heat engines, can be realized if the input incoherent radiation is broad enough to excite all the nearly degenerate energy transitions, thereby producing interference between them.

Implementing noise-induced Fano coherence in a laboratory requires physical platforms that offer suitable energy level structures, controllable incoherent excitation, and tunable dynamical parameters. Three major classes of physical systems satisfy these conditions: (i) atomic ensembles such as hot or cold clouds of Rb-87, (ii) solid-state quantum dots, and (iii) superconducting artificial atoms based on circuit quantum electrodynamics (circuit QED).

Both atomic vapors and cold-atom systems provide clean dipole-allowed V-type transitions and well-characterised decoherence channels. These platforms enable proof-of-principle explorations, with a tunable excited state splitting (e.g., via magnetic field) and compatibility with broadband incoherent pumps closely matching the theoretical framework developed here.\\
Superconducting circuits provide another promising platform, thanks to high control over energy transition frequencies, decay rates, dipole orientations, and engineered reservoirs. Experiments have already demonstrated that multi-level superconducting qubits can reproduce atomic three-level phenomena such as electromagnetically induced transparency based on the Autler-Townes splitting with large fidelity~\cite{Hoi}. The tunability of superconducting qubits makes them ideal for implementing V-type systems with adjustable decay asymmetry, and for applying tailored broadband or thermal-like microwave fields. Hence, circuit QED can be an excellent candidate for testing quantum coherence generation due to incoherent driving in regimes difficult to access in atomic ensembles or solid-state devices.\\ 
Finally, semiconductor quantum dots are highly promising as photo-conversion technologies for photovoltaic applications. Indeed, their engineered band structure and light-matter couplings align them well with the V-type photocell architecture proposed by Svidzinsky et al.~\cite{Svidzinsky}. In these systems, conduction band states naturally form the excited manifold while the valence band plays the role of the ground state. Incoherent sunlight or thermal radiation acts as the pump. Several theoretical works have predicted that, under these conditions, coherence-assisted suppression of radiative recombination will occur, as well as an enhancement of photocurrent and open-circuit voltage~\cite{Dorfman,Svidzinsky,Scully}.

In the rest of this section, we will discuss in more detail how a platform based on Rubidium atoms can host incoherently driven V-type systems. We will examine the advantages of this platform, as well as the challenges associated with observing Fano coherence in realistic experimental conditions.

\subsection{Ensemble of Rubidium atoms}\label{subsec:3.1}

Despite extensive research done on the topic, in our knowledge, an experiment clearly proving the generation of Fano coherences by incoherent radiation is still missing.

Atomic systems represent the natural experimental platform for a proof-of-principle demonstration of noise-induced Fano coherence. They offer several intrinsic advantages: their electronic structure is precisely known, and selection rules are unambiguous. Furthermore, decoherence channels and key parameters such as transition dipoles, linewidths, and level splittings can be characterised and tuned with great accuracy. For an atomic species to serve as a suitable candidate for Fano coherence generation, its level structure must contain: (i) two excited levels connected to a common ground level via electric-dipole-allowed transitions; (ii) excited levels whose energy splitting can be tuned over a broad range, without introducing additional couplings; (iii) transition frequencies compatible with available broadband incoherent radiations.

A particularly useful feature to generate Fano coherences is the ability to implement the excited manifold using magnetic sublevels, which enables continuous and well-defined tuning of the energy splitting $\Delta$ by varying an external magnetic field. This controllability is essential for accessing both the underdamping and overdamping dynamical regimes predicted by the theory. Likewise, the weak- and strong-pumping regimes (defined by the mean photon number $\bar{n}$) can be explored by adjusting the intensity of the broadband incoherent radiation, for example by varying the power of a broadband laser source.

Isotopes with rich hyperfine and Zeeman manifolds, such as alkali and alkaline earth atoms, naturally fulfil the criteria described above. Their well-documented spectroscopic properties and the technological maturity of laser systems at their transition frequencies further simplify implementation. Previous proposals, such as those by Dodin \emph{et al.}~\cite{Dodin-Brumer} and subsequently by Koyu \emph{et al.}~\cite{Koyu-Dodin}, suggested experiments with beams of Calcium atoms excited by a broadband polarized laser, within a uniform magnetic field. Similarly, in Ref.~\cite{Koyu} the authors propose an experimental scenario for measuring steady-state noise-induced Fano coherences in a $\Lambda$-type three-level system, using metastable He($2^3S_1$) atoms. Additionally, recent experiments with a magneto-optical trap of Rubidium atoms~\cite{Han-Lee} have shown an increase in the beat amplitude, which is the oscillatory modulation of the emitted radiation intensity, originating from the presence of quantum coherence between the excited levels of the system. This effect proves the generation of vacuum-induced quantum coherences, which arise from the coupling of discrete atomic states with vacuum modes of the electromagnetic field.

\subsection{Implementation of V-type system in $^{87}$Rb hyperfine structure}\label{subsec:3.2}

An atomic platform based on Rubidium-87 ($^{87}$Rb) atoms is a suitable candidate for a proof-of-principle experiment to detect noise-induced Fano coherence, thanks to its accessible optical transitions and versatile hyperfine structure. The hyperfine Zeeman sublevels of the $5^{2}S_{1/2}$ ground state manifold, and the $5^{2}P_{1/2}$ and $5^{2}P_{3/2}$ excited states manifolds can host multiple realizations of a V-type configuration.
These levels exhibit linear Zeeman shifts at low magnetic fields, which enables fine control over the excited state energy splitting. Such a requirement is central to exploring the different dynamical regimes of the V-system. 
It is also needed that the selected hyperfine transitions are spectrally isolated from the neighbouring manifolds, such that no unwanted transitions are driven. This condition sets a constraint on the bandwidth of the broadband radiation that is used as the incoherent pump. In particular, the input radiation must be wide enough in the frequency domain in order to behave as an incoherent source, yet narrow enough to avoid coupling to other hyperfine or fine structure levels. A particularly convenient choice to fulfil the requirements above is the $5^{2}S_{1/2}\leftrightarrow5^{2}P_{1/2}$ ($D_1$) transition, where the hyperfine manifolds of the ground state ($F=1$ and $F=2$) are separated by approximately $6.83$ GHz, while the two excited level manifolds ($F'=1$ and $F'=2$) are separated by $814$ MHz~\cite{Steck,Steck:Rb87}. Notably, both the excited and ground manifolds can be resolved even in hot vapor, despite the presence of the Doppler broadening due to atomic motion. Finally, the D1 transition can be driven using standard diode lasers at $\lambda = 795$ nm and the decay rate of such transition is approximately equal to $2\pi \times 5.75$ MHz. 

\begin{figure}[t]
\centering
\includegraphics[width=0.55\textwidth]{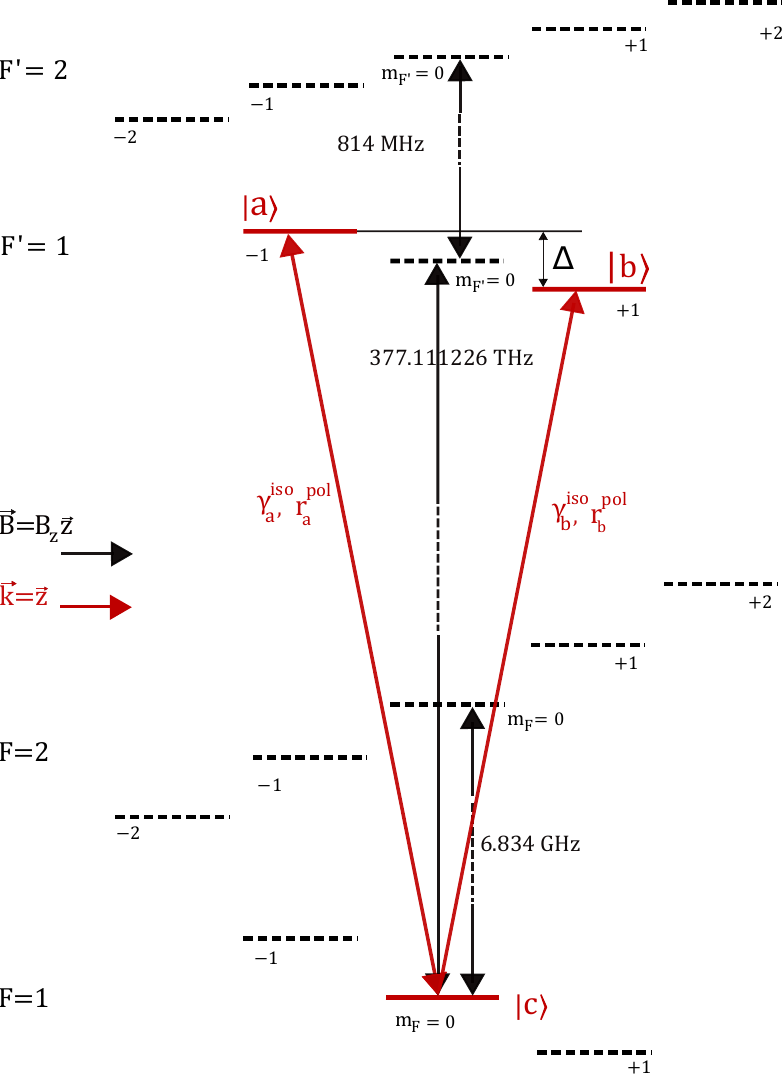}
\caption{V-type three level system within the D1 line transition of $^{87}$Rb. The ground state $|c\rangle$ corresponds to the hyperfine magnetic sublevel $|F=1 \,, m_F=0\rangle$, while the two excited states $|a\rangle$ and $|b\rangle$ are represented by the levels $|F'=1 \,, m_{F'}=-1\rangle$ and $|F'=1 \,, m_{F'}=+1\rangle$, respectively. The red arrows denote the radiation processes involved: spontaneous emission from levels $|a\rangle $ and $|b\rangle $ to level $|c\rangle$ at rates $\gamma_{a}^{\rm iso}=\gamma_{b}^{\rm iso}$ respectively; incoherent pumping and stimulated emission involving transitions $|a\rangle \leftrightarrow |c\rangle$ and $|b\rangle \leftrightarrow |c\rangle$ at rates $r_a^{\rm pol}=r_b^{\rm pol}$, respectively. The parameter $\Delta$ is the energy splitting between the excited levels: $\Delta=\omega_{ac}-\omega_{bc}$. In the figure, the wavevector ${\bf k}$ of the radiation and the uniform magnetic field vector ${\bf B}$ are reported, as well as the frequency difference between the hyperfine manifolds of the $D1$ line.} 
\label{fig:4}
\end{figure}

Among the four allowed hyperfine transitions connecting the $F=1,2$ ground manifolds to the $F'=1,2$ excited manifolds, it is worth considering the transition $F=1\rightarrow F'=1$ as it contains the smallest number of magnetic sublevels, namely $2F+1=3$ in each manifold~\cite{Steck,Steck:Rb87}.
Within this transition, the implementation of the V-scheme is illustrated in Fig.~\ref{fig:4}. The ground state $|c\rangle$ corresponds to the atomic level $|F=1 \,, m_F=0\rangle$, while the two excited states $|a\rangle$ and $|b\rangle$ are represented by levels $|F'=1 \,, m_{F'}=-1\rangle$ and $|F'=1 \,, m_{F'}=+1\rangle$, respectively.
The transitions $|a\rangle \leftrightarrow |c\rangle$ and $|b\rangle \leftrightarrow |c\rangle$ have the same spontaneous decay rate (i.e., $\gamma_a^{\rm iso}=\gamma_b^{\rm iso}$), thus the V-system decays symmetrically. Furthermore, they are driven by circularly polarized light in the x-y plane since they adhere to the selection rules defined by the difference $\Delta m_F = -1$ and $\Delta m_F = +1$, respectively~\cite{Steck}. As a consequence, the corresponding dipole moment vectors $\bm{\mu}_{ac}$ and $\bm{\mu}_{bc}$ are of the form derived in Subsec.~\ref{subsec:3} and orthogonal, yielding $p=0$. This means that no quantum interference occurs between the spontaneous emission channels, while interference in absorption and stimulated emission is allowed if the pumping field has the appropriate polarization. To drive both transitions simultaneously, the broadband radiation must have a linear polarization in the transverse x-y plane; for consistency with the theoretical analysis of Subsec.~\ref{subsec:3}, we take the linear polarization along the x-axis, $\bm{\epsilon}_x$ (see Eq.~\eqref{eq:1.42a}). The corresponding pumping rates of the two transitions are $r_a^{\rm pol}=r_b^{\rm pol}$. Under these conditions, the master equation in Eqs.~\eqref{eq:1.47a}-\eqref{eq:1.47b} accurately describes the dynamics of the proposed atomic system, provided that the approximations used for its derivation remain valid. 
For this reason, let us examine each of these approximations and identify the corresponding validity regime for the $^{87}$Rb D1 platform.
\begin{itemize}
\item {\bf Validity of the dipole approximation:}\\
For the D1 transition of $^{87}$Rb ($\lambda=795$ nm), the spatial extent of the electronic wave-function is on the order of the Bohr radius $a_0\approx0.053$ nm. This means that the atom is more than $10^{3}-10^{4}$ times smaller than the optical wavelength \cite{Steck:Rb87}. Therefore, the optical field can be considered effectively uniform across the atomic extension, and the interaction Hamiltonian is well-described by the electric dipole terms in all operating regimes relevant to our implementation. 

\item {\bf Validity of the rotating-wave approximation (RWA):}\\
For the D1 transition of $^{87}$Rb, the atomic transition frequencies $\omega_{ac}\simeq \omega_{bc}\approx 2\pi\times 400$ THz are several orders of magnitude larger than both the spontaneous decay rate ($\approx 2\pi \times 5.75$ MHz) and incoherent pumping rates ($\approx 2\pi$ GHz in the strong pumping regime), which set the effective timescale of the system-reservoir coupling. Furthermore, when the broadband radiation is spectrally centred near the atomic transition, such that the dominant field modes satisfy $|\omega_{lc}-\nu_{{\bf k},\bm{\epsilon}_x}|\ll\omega_{lc}+\nu_{{\bf k},\bm{\epsilon}_x}$, a clear separation of timescales between the fast oscillations of the counter-rotating terms and the slow system's dynamics is ensured. Consequently, non-resonant processes associated with counter-rotating terms are strongly suppressed, and the RWA well describes the interaction of the $^{87}$Rb D$_1$ hyperfine V-system with a broadband incoherent radiation field.

\item {\bf Validity of the Born-Markov approximation:}\\
For the implementation proposal, we consider a broadband laser field employed as an incoherent radiation source, characterised by a spectral bandwidth $\Delta\nu_{\rm laser}$. As discussed in Subsec.~\ref{subsubsec:1}, the correlation time of the input radiation $\tau_R$ can be estimated as $\tau_R \simeq 1/\Delta\nu_{\rm laser}$.

On the other hand, the characteristic timescale $\tau_S$ of the atomic system built over the $^{87}$Rb D$1$ line is determined by the rate of spontaneous emission from the excited manifold, with an average decay rate $\bar{\gamma} \approx 2\pi \times 5.75~\mathrm{MHz}$ that corresponds to $\tau_S = 2\pi/\bar{\gamma}\approx 200~\mathrm{ns}$. The Markov approximation requires a clear separation of timescales, $\tau_R \ll \tau_S$, which translates into the condition $\Delta\nu_{\rm laser} \gg \bar{\gamma}$.

Therefore, spectral bandwidths in the range of several tens to a few hundreds of MHz ensure that the reservoir's correlation time $\tau_R$ is much shorter than the relaxation time $\tau_S$ of the system, justifying the Markov approximation for the present setup. At the same time, the bandwidth must remain sufficiently narrow to avoid significant excitation of nearby hyperfine levels. In particular, the excited state manifolds $F'=1$ and $F'=2$ of the D$1$ line are separated by approximately $814~\mathrm{MHz}$. Choosing $50~\mathrm{MHz}\lesssim\Delta\nu_{\rm laser} \lesssim 500~\mathrm{MHz}$ guarantees spectral selectivity, while guaranteeing the Born-Markov approximation. This is further supported by the fact that the two transitions $F=1 \rightarrow F'=1$ and $F=1 \rightarrow F'=2$ are spectrally resolvable even at room temperature, where Doppler broadening is on the order of $500~\mathrm{MHz}$. This implies that a radiation with this bandwidth does not excite the $F=1 \rightarrow F'=2$ hyperfine transition.
\end{itemize}

\subsubsection{Theoretical predictions}\label{subsec:3.2.1}

The V-type three-level system, implemented in the hyperfine structure of $^{87}$Rb atoms, interacts with a broadband laser, with the wavevector ${\bf k}$ along the z-axis and linear polarization along the x-axis. The dynamical regimes with anisotropic polarized radiation have been analysed in Subsec.~\ref{subsec:2.3}, observing that Fano coherence reaches a non-zero stationary value in both the underdamping and overdamping regimes. However, in the overdamping regime, the quantum coherence evolves monotonically without oscillations over time, which allows that the stationary value of coherence is optimized for efficient detection

We now analyse the V-system following the approach introduced in Subsec.~\ref{subsec:strong-pumping regime}, investigating the conditions under which stationary Fano coherence $|\rho_{ab}|_{\rm steady}$ can be maximized in the implementation with $^{87}$Rb atoms. For this purpose, we vary the parameters $\Delta/\bar{\gamma}$ and $\bar{n}$ within the overdamping regime. 
\begin{figure}[t]
\centering
\includegraphics[width=0.8\textwidth]{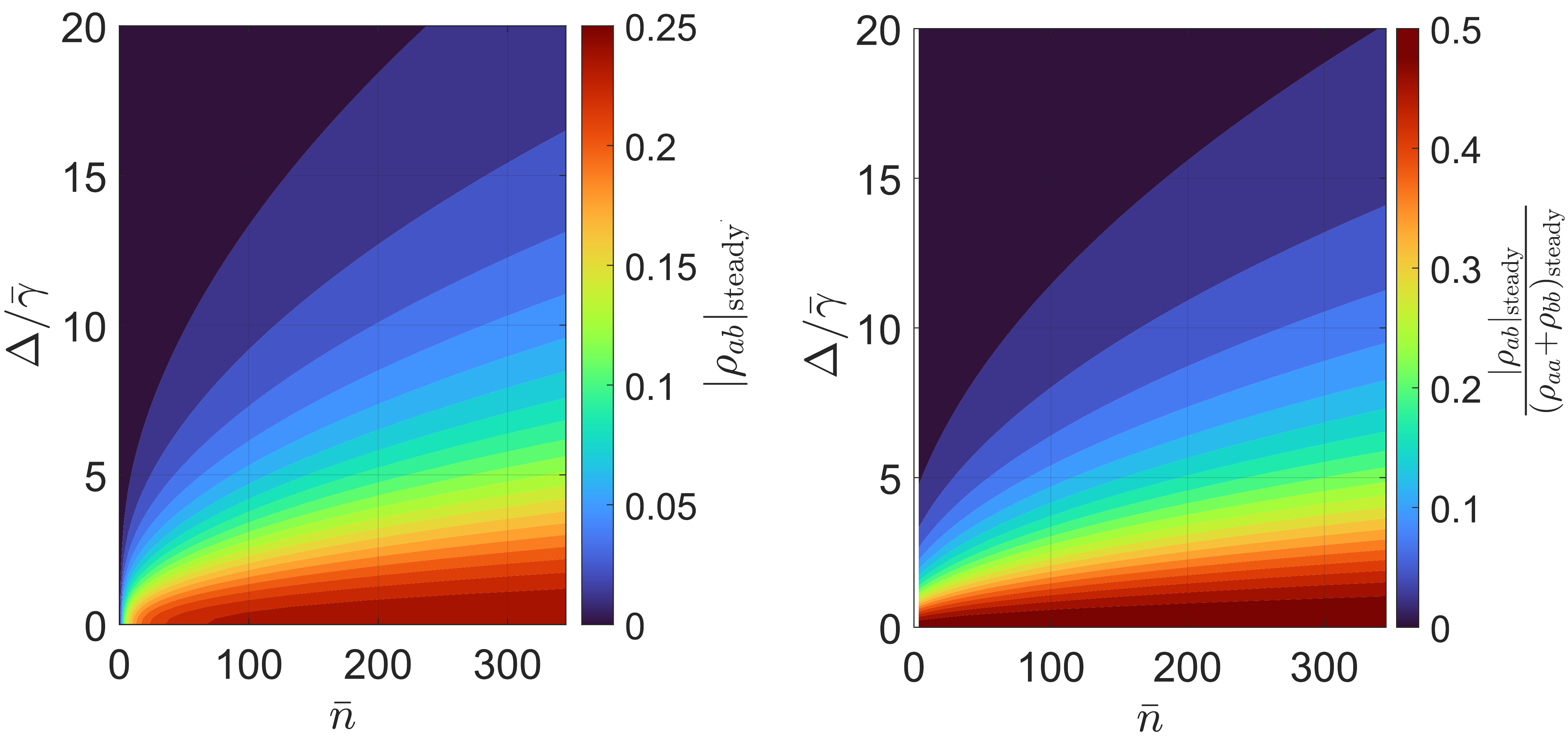}
\caption{(a) Stationary magnitude of the quantum coherence $|\rho_{ab}|$ as a function of the mean photon number $\bar{n}$ and of the normalized excited state splitting $\Delta/\bar{\gamma}$.
(b) Stationary coherence normalized with respect to the excited state populations, as a function of $\bar{n}$ and $\Delta/\bar{\gamma}$. The V-type three-level system is driven by a linearly polarized radiation field along the $x$-axis.} 
\label{fig:5}
\end{figure}
Fig.~\ref{fig:5} displays the stationary values of $|\rho_{ab}|$ for different combinations of $\Delta/\bar{\gamma}$ and $\bar{n}$. The maximum value of the mean photon number considered in this analysis is $\bar{n}_{\rm max}=345$, beyond which the radiation intensity exceeds the saturation intensity of the atomic transition. Maintaining the intensity of the input radiation below saturation is essential, since saturation of the transitions $|a\rangle\leftrightarrow|c\rangle$ and $|b\rangle\leftrightarrow|c\rangle$ leads to a reduced absorption and ultimately to the degradation of noise-induced Fano coherences.

From Fig.~\ref{fig:5}(a), it is evident that a large excited state energy splitting is detrimental to the formation of stationary Fano coherence, identifying the regime $0 < \Delta/\bar{\gamma} < 1$ as the most favourable. Moreover, increasing the mean photon number $\bar{n}$ enhances the intensity of quantum coherence at the steady-state. Therefore, in order to observe and detect stationary Fano coherences in experimental implementation, one needs to combine a small excited state energy splitting and a moderately strong incoherent pumping.
Fig.~\ref{fig:5}(b) shows the behaviour of the coherence ratio 
$\frac{|\rho_{ab}|_{\rm steady}}{(\rho_{aa}+\rho_{bb})_{\rm steady}}$ as a function of $\Delta/\bar{\gamma}$ and $\bar{n}$. Such quantity is a convenient measure of the coherence strength with respect to the excited state populations, and can be evaluated using observables such as fluorescence emission intensities~\cite{Dodin-Tscherbul,Tscherbul-Brumer2,Koyu-Dodin}. The maximum attainable value of the ratio is $1/2$, which also occurs in this case under a small excited state splitting and strong, yet unsaturated, pumping.

%%%%%%%%%%%%%%%%%%%%%%%%%%%%%%%%
\section{Conclusions}\label{sec4}

In this work, we have developed a first-principle quantum model to describe the generation and persistence of Fano coherence in a V-type three-level quantum system driven by polarized incoherent radiation. By deriving the Bloch-Redfield master equation without imposing full secular approximation, we have confirmed that quantum interference between the nearly degenerate excited states of the three-level system can be stationary under realistic conditions. Our analysis identifies the dynamical regimes---spanned by the ratio between the excited state energy splitting and the spontaneous decay rate due to the interaction with the field---where the generation of Fano coherence is maximized. In particular, a small energy splitting combined with a moderately strong incoherent pumping allow for the stationarity of Fano coherence in a robust manner, namely even in the presence of spontaneous emission paths.

We complemented the theoretical predictions with a discussion about the experimental feasibility of the model, focusing on atomic ensembles of Rubidium-87 that offer controllable level structures and compatibility with broadband incoherent sources. Our discussion suggests that noise-induced quantum coherence can be harnessed in practical platforms, paving the way for applications in quantum-enhanced energy conversion and information storage.

\subsection{Outlooks}\label{subsec:4.1}

Several research directions can emerge from this work:

\begin{enumerate}
    \item[i)] {\it Energy conversion efficiency:}\\
    A rigorous quantification of how stationary Fano coherence influences energy conversion in quantum heat engines and photocells remains an open challenge. Future investigations should explore the dependence of conversion efficiency on the degree of anisotropy and polarization of an incoherent radiation field in input, as well as on the excited state splitting and pumping intensity. Particular attention should be given to identifying regimes where Fano coherence acts as a thermodynamic resource, potentially reducing entropy production and enhancing extractable work.
         
    \item[ii)] {\it Experimental verification and control strategies:} \\
    The design of experiments capable of detecting and controlling Fano coherence under realistic conditions is a crucial step to be still addressed. Atomic platforms, such as an ensemble of Rubidium atoms, offer a natural proof-of-principle implementation thanks to their well-characterised level structure and tunability. Beyond atomic systems, superconducting circuits and semiconductor quantum dots could be promising alternatives due to their scalability and controllable parameters. Exploring these platforms will enable the integration of coherence-assisted mechanisms into energy-harvesting devices and quantum thermodynamic cycles.

    \item[iii)] {\it Exploiting Fano coherence as a resource in non-equilibrium processes:} \\ 
    Further research should address the question how Fano coherence can be systematically exploited as a resource in non-equilibrium processes and energy conversion schemes, through coherent or incoherent interaction with an external load. From this perspective, it would be particularly useful to understand how Fano coherence can be converted into a photocurrent within the load, which can then be stored or used for powering devices.
\end{enumerate}

%%%%%%%%%%%%%%%%%%%%%%%%%%%%%%%%
\subsection*{Acknowledgments}

The Authors acknowledge very useful discussions with Natalia Bruno and Chiara Mazzinghi. This work has been financial supported by the PNRR MUR project PE0000023-NQSTI funded by the European Union---Next Generation EU, and from ASI and CNR under the Joint Project ``Laboratori congiunti ASI-CNR nel settore delle Quantum Technologies (QASINO)'' (Accordo Attuativo n. 2023-47-HH.0).

%%%%%%%%%%%%%%%%%%%%%%%%%%%%%%%%
\subsection*{Authors' contributions}

L.D. and S.G. performed the theoretical calculation and numerical simulations with the inputs from F.S.C.. All authors conceived the project, carried out the analysis of the results, contributed to the discussion, and the writing of the manuscript.

%%%%%%%%%%%%%%%%%%%%%%%%%%%%%%%%
\bibliography{sn-bibliography}

\end{document}